\documentclass[prd,twocolumn,nopacs,floatfix,amsmath,nofootinbib,amssymb,preprintnumbers]{revtex4-2}
\usepackage{graphicx,subcaption,color,dcolumn,booktabs,diagbox}
\usepackage{txfonts}
\usepackage{slashed}
\usepackage{epstopdf}
\usepackage{multirow}
\usepackage{bm}
\usepackage{braket}
\usepackage{pifont}
\usepackage{adjustbox}
\usepackage{rotating}
\usepackage{verbatim}
\usepackage{appendix}
\usepackage{graphicx,xcolor,dcolumn,booktabs}
\usepackage[colorlinks,
            citecolor=blue,
            anchorcolor=red,
            menucolor=red,
            linkcolor=red,
            filecolor=red,
            runcolor=red,
            urlcolor=blue,
            frenchlinks=true]{hyperref}
\begin{document}
\title{Prospects for observing the missing $2D$ and $1F$ charmonium states around 4 GeV}

\author{Cheng-Xi Liu$^{1,2,3,4}$}\email{liuchx2025@lzu.edu.cn}
\author{Zi-Long Man$^{1,2,3,4}$}\email{manzl@lzu.edu.cn}
\author{Tian-Le Gao$^{1,2,3,4}$}\email{gaotl2025@lzu.edu.cn}
\author{Xiang Liu$^{1,2,3,4}$}
\email{xiangliu@lzu.edu.cn}
\affiliation{
$^1$School of Physical Science and Technology, Lanzhou University, Lanzhou 730000, China\\
$^2$Lanzhou Center for Theoretical Physics, Key Laboratory of Theoretical Physics of Gansu Province, Key Laboratory of Quantum Theory and Applications of MoE, Gansu Provincial Research Center for Basic Disciplines of Quantum Physics, Lanzhou University, Lanzhou 730000, China\\
$^3$MoE Frontiers Science Center for Rare Isotopes, Lanzhou University, Lanzhou 730000, China\\
$^4$Research Center for Hadron and CSR Physics, Lanzhou University and Institute of Modern Physics of CAS, Lanzhou 730000, China}

\begin{abstract}

Our understanding of high-lying states within the charmonium family remains incomplete, particularly in light of recent observations of charmonium states at energies around 4 GeV. In this study, we investigate the spectroscopic properties of several high-lying charmonia, focusing on the $2D$ and $1F$ states. A mass spectrum analysis is conducted, incorporating the unquenched effects. We then present a detailed study of the strong decay properties, including partial decay widths for two-body strong decays permitted by the Okubo-Zweig-Iizuka (OZI) rule. Additionally, we explore the primary radiative decay channels associated with these states. Finally, we discuss the radiative transitions of the $2D$ and $1F$ states via $e^+e^-$ annihilation.  Theoretical predictions provided here aim to guide future experimental searches for high-lying charmonium states at facilities such as BESIII, Belle II, LHCb, and the future STCF.

\end{abstract}

\maketitle

\section{Introduction}

2024 marks the 50th anniversary of the discovery of the $J/\psi$ particle, the first charmonium state observed in 1974 \cite{E598:1974sol, SLAC-SP-017:1974ind}. Since then, over a dozen charmonium states have been identified, including the $\psi(3686)$ \cite{Abrams:1974yy}, $\eta_{c2}(1S)$ \cite{Himel:1980dj}, $\eta_{c2}(2S)$ \cite{Edwards:1982fif}, $\psi(4040)$ \cite{Goldhaber:1977qn}, $\psi(4415)$ \cite{Siegrist:1976br}, $h_{c2}(1P)$ \cite{Armstrong:1992ae}, $\chi_{c0}(1P)$ \cite{Biddick:1977sv}, $\chi_{c1}(1P)$ \cite{Tanenbaum:1975ef}, $\chi_{c2}(1P)$ \cite{Whitaker:1976hb}, and $\psi(3770)$ \cite{Rapidis:1977cv}, $\psi(4160)$ \cite{DASP:1978dns}.
The discovery of these low-lying charmonium states motivated the development of the Cornell potential model \cite{Eichten:1974af, Eichten:1978tg}, which provided a quantitative framework for characterizing hadron spectroscopy.

However, the charmonium family is still not yet fully established. Entering the 21st century, the discovery of a series of $XYZ$ charmonium-like states (see reviews \cite{Liu:2009qhy, Liu:2013waa, Chen:2016qju,Guo:2017jvc,Brambilla:2019esw,Chen:2022asf,Deng:2023mza, Chen:2013cpa} for details) has stimulated extensive discussions on identifying new charmonia again. With accumulating experimental data, the $\psi_2(3823)$-a promising candidate for the $1D$-wave charmonium state-was first observed in the $B^\pm\to \chi_{c1}\gamma K^\pm$ decay \cite{Belle:2013ewt}, while the $\psi_3(3842)$ was found in the $D\bar{D}$ invariant mass spectrum of the $pp\to D\bar{D}+\mathrm{anything}$ process \cite{LHCb:2019lnr}. These findings complete the $1D$-wave charmonium triplet. Nevertheless, the $1^3D_2(2^{-+})$ charmonium state remains unobserved.  
 
$X(3872)$ \cite{Belle:2003nnu}, $X(3915)$ \cite{Belle:2009and}, and $Z(3930)$ \cite{Belle:2005rte} may constitute a $2P$-wave charmonium triplet. This suggests that unquenched effects are essential in hadron spectroscopy studies, especially for decoding the properties of high-lying charmonium states. In these states, open-charm decay channels become accessible, meaning the coupling between bare $c\bar{c}$ configurations and their open-charm channels can no longer be neglected. This insight is not confined to charmonium (see the relevant discussion for the $D_{s0}(2317)$ \cite{Luo:2021dvj}, $D_{s1}(2460)$ \cite{vanBeveren:2003jv}, and $\Lambda_{c2}(2940)$ \cite{Luo:2019qkm}). In addressing the ``$Y$ problem", our analysis revealed an unquenched charmonium mass spectrum spanning 4.0-4.5 GeV \cite{Wang:2019mhs}. This framework not only provides natural assignments for the $Y(4230)$ state, but also predicts the existence of $\psi(4380)$ and $\psi(4500)$ charmonia. These predictions align with the recent high-precision BESIII data \cite{BESIII:2022joj,BESIII:2018iea}.

Through collaborative efforts between experimentalists and theorists, the catalog of established charmonium states continues to expand. However, significant gaps remain in the mass spectrum of higher partial-wave charmonium states near 4 GeV. For instance, while the $Y(4160)$ has been identified as a $2^3D_1$ charmonium state, its three expected $2D$-wave partners have yet to be observed. Furthermore, following the established pattern of sequential excitation, the anticipated $1F$-wave charmonium states-which should appear after the $D$-wave states-are currently absent from current observations.

We now turn to ongoing experiments relevant to charmonium studies. As the sole experiment operating in the charm-tau energy region, BESIII is one of the central to charmonium research. Over the past decade, BESIII has accumulated substantial $e^+e^-$ collision data at multiple energy points above 4 GeV. This work focuses on recent experimental progress to evaluate the discovery potential for $2D$ and $1F$ charmonium states. While BESIII provides crucial data, complementary opportunities exist at LHCb and Belle II, where $B$ meson weak decays offer alternative production mechanisms for these charmonium states. We examine these various approaches. Our primary objective is to advance the spectroscopy of $2D$ and $1F$ charmonium states, addressing current gaps in our understanding of higher orbital excitations. 

This paper is organized as follows. In Sec. \ref{sec2}, we briefly review the present status of charmonium family and previous studies on charmonium states. Then, in Sec. \ref{sec3}, we introduce the phenomenological models employed, including the modified Godfrey-Isgur (MGI) model for the mass spectrum, the quark pair creation model for Okubo-Zweig-Iizuka (OZI)-allowed strong decays, the constituent quark model for electromagnetic transitions. Our numerical results and corresponding analysis are presented in Sec. \ref{sec4}. And in Sec. \ref{sec5} discusses the production of these charmonium states using the hadronic loop mechanism. Finally, we summarize and explore how the $2D$ and $1F$ may be detected.

\section{The present status of charmonium} \label{sec2}

The observation of low-lying charmonia made the characterization of the charmonium spectrum a central topic in the late twentieth century. A key development was the introduction of the Cornell potential for $c\bar{c}$ interactions by the Cornell group \cite{Eichten:1974af,Eichten:1978tg}, which enabled a quantitative description of hadron spectroscopy. Building upon this foundation, various potential models were proposed, including the well-known Godfrey-Isgur (GI) model \cite{Godfrey:1985xj}. It should be noted that the potential models mentioned above are typical quenched models.

\begin{figure*}[htbp]
	\centering
	\includegraphics[width=1\textwidth]{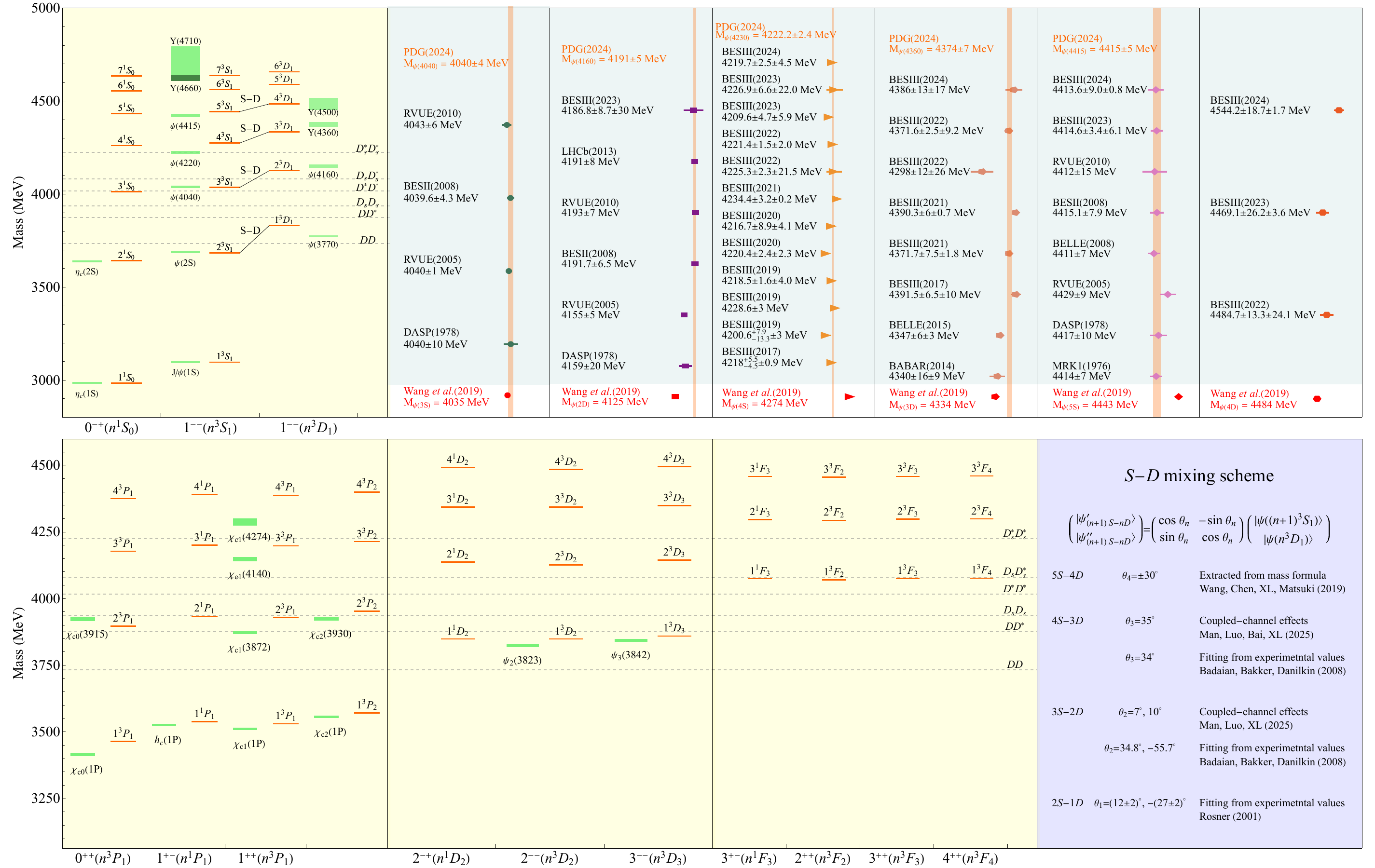}\\[2em]       
 \captionsetup{justification=raggedright, singlelinecheck=false}
\caption{The mass spectrum of charmonium family, the experimental data and $S$-$D$ mixing scheme of vector charmonia. }\label{mass spectrum} 
\end{figure*}

Under the quenched picture, the experimental assignments for these charmonium states are as follows:
\begin{itemize}
    \item For the $S$-wave states: $\eta_c(1S)$ $(1^1S_0)$ \cite{Himel:1980dj}, $\eta_c(2S)$ \cite{Edwards:1982fif}, $J/\psi$ ($1^3S_1$) \cite{E598:1974sol, SLAC-SP-017:1974ind}, $\psi(3686)$ ($2^3S_1$) \cite{Abrams:1974yy}, $\psi(4040)$ ($3^3S_1$) \cite{Goldhaber:1977qn}, and $\psi(4415)$ ($4^3S_1$) \cite{Siegrist:1976br};

 \item For the $P$-wave states: $h_c(3525)$ $(1^1P_1)$ \cite{Armstrong:1992ae}, $\chi_{c0}(3415)$ $(1^3P_0)$ \cite{Biddick:1977sv}, $\chi_{c1}(3510)$ $(1^3P_1)$ \cite{Tanenbaum:1975ef}, and $\chi_{c2}(3556)$ $(1^3P_2)$ \cite{Whitaker:1976hb}.

 \item For the $D$-wave states: $\psi(3770)$ $(1^3D_1)$ \cite{Rapidis:1977cv} and $\psi(4160)$ $(2^3D_1)$ \cite{DASP:1978dns}.

\end{itemize}
These assignments were applied in concrete studies, including experimental analyses. However, for the assignment of high-lying charmonia, the limitations of the quenched model become evident.  In Fig.~\ref{mass spectrum}, the relative positions of the charmonium states are illustrated alongside experimental data.

The discovery since 2003 of multiple charmonium-like $XYZ$ states has presented a major challenge to these models \cite{Liu:2009qhy, Liu:2013waa, Chen:2016qju,Guo:2017jvc,Brambilla:2019esw,Chen:2022asf,Deng:2023mza}. The first and seminal example, the $X(3872)$ \cite{Belle:2003nnu}, exhibits a mass lower than the $\chi_{c1}(2P)$ state predicted by quenched quark models. Subsequently, the Belle Collaboration reported the $Z(3930)$ in the $D\bar{D}$ invariant mass spectrum \cite{Belle:2005rte}, and the $X(3915)$ in the $\gamma\gamma \to J/\psi\omega$ process \cite{Belle:2009and}. Within the quenched quark model, the $Z(3930)$ is often interpreted as the $\chi_{c2}(2P)$ state, whereas the assignment of the $X(3915)$ as the $\chi_{c0}(2P)$ remains controversial.
 
 The Lanzhou group proposed that the $X(3915)$ remains a strong candidate for the charmonium $\chi_{c0}(2P)$ state \cite{Liu:2009fe}. However, this assignment raises several questions \cite{Guo:2012tv}. Specifically, the mass splitting between the $X(3915)$ and $Z(3930)$ is smaller than the $\chi_{c0}(2P)$ and $\chi_{c2}(2P)$ states expected from the quenched quark model. To address this discrepancy, the Lanzhou group conducted further studies \cite{Duan:2020tsx} and demonstrated that the small mass gap can be explained when coupled-channel effects are incorporated into the charmonium spectrum. With such effects included, the $X(3915)$, $X(3872)$, and $Z(3930)$ can be assigned as the $\chi_{c0}(2P)$, $\chi_{c1}(2P)$, and $\chi_{c2}(2P)$ states, respectively. Additionally, the $\chi_{c1}(4274)$ state has been observed in the $B^+ \to J/\psi \phi K^+$ process \cite{LHCb:2016axx, CDF:2011pep}. The study of the $2P$ charmonium states illustrates that unquenched effects are essential for describing higher radial and orbital excitations. Consequently, the Lanzhou group employed an unquenched model to describe the $3P$, $4P$, and $5P$ charmonium states, identifying the $\chi_{c1}(4270)$ as a candidate for the $\chi_{c1}(3P)$ state \cite{Duan:2021alw}.

The $^3D_1$ and $^3S_1$ states share the same $J^{PC}$ quantum numbers, implying an $S$–$D$ mixing scheme. This mixing framework was originally proposed to resolve two important puzzles: the absence of the $\psi(3686) \to \rho\pi$ decay \cite{Franklin:1983ve, BES:1998dmn} and the anomalous ratio $\Gamma(\psi(3770) \to \gamma\chi_{c2}) / \Gamma(\psi(3770) \to \gamma\chi_{c0})$ \cite{Kwong:1988ae}. The $S$–$D$ mixing scheme successfully accounts for both phenomena \cite{Rosner:2001nm}. Since the masses of $\psi(3686)$ and $\psi(3770)$ lie near the $D\bar{D}$ threshold, strong coupling to charmed meson pairs induces $S$–$D$ mixing, with the mixing angle constrained by their di-electron widths. This framework results in an enhanced $\Gamma(\psi(3770) \to \gamma\chi_{c2})$ and a suppressed $\psi(3686) \to \rho\pi$ process, causing the latter to instead appear in the $\psi(3770)$ decay width.

Although the vector states $\psi(4040)$ and $\psi(4160)$ are well described in the quenched framework, recent experimental results present a new puzzle. A key issue is the mass of $\psi(4160)$: it has shifted upward from the initial measurement of $4160$ MeV \cite{DASP:1978dns} to a recent value of $4190$ MeV \cite{BES:2007zwq, LHCb:2013ywr}. As $\psi(4160)$ is often used as a scaling point in phenomenological models, this shift challenges the established charmonium spectrum. The Lanzhou group reevaluated the dimuon spectrum by incorporating contributions from higher charmonia \cite{Peng:2024blp}. Their best fit supports a $\psi(4160)$ mass of about 4145 MeV. Furthermore, Calculations indicate that for the mass exceeding 4170 MeV, the experimental data cannot be well fitted, suggesting a relatively low mass for this state. In addition, the nearly equal di-electron widths of $\psi(4040)$ and $\psi(4160)$ \cite{DASP:1978dns, Seth:2004py} indicate $S$–$D$ wave mixing. The $3S$–$2D$ mixing scheme has been studied in Refs. \cite{Badalian:2008dv, Wang:2022jxj, Zhao:2023hxc, Man:2025vmm}. Among these, the analysis by the Lanzhou group derived mixing angles and masses consistent with a lower mass for $\psi(4160)$ \cite{Peng:2024blp}.

Predictions from unquenched models suggest the presence of more resonances between $\psi(4160)$ and $\psi(4415)$. In 2013, the BESIII Collaboration reported a narrow structure around 4.2 GeV \cite{BESIII:2013ouc, Yuan:2013uta}, which can be identified as $\psi(4S)$ \cite{Chen:2011kc, Chen:2014sra}. Later, BESIII reported results from $e^+e^-\to \pi^+\pi^-J/\psi$ \cite{BESIII:2016bnd} and $e^+ e^- \to \pi^+ \pi^- h_c$ \cite{BESIII:2016adj}, revealing broad asymmetric structures in the $4.2-4.4$ GeV region, including three charmonium-like states: $Y(4220)$ and $Y(4320)$ \cite{BESIII:2016bnd}, and $Y(4390)$ \cite{BESIII:2016adj}. In addition, the BaBar and Belle collaborations reported charmonium-like states $Y(4008)$, $Y(4260)$, and $Y(4360)$ \cite{BaBar:2005hhc, Belle:2007dxy, Belle:2013yex, BaBar:2006ait, Belle:2007umv, Belle:2014wyt}. The Lanzhou group analyzed the data using the Fano-like interference effect \cite{Chen:2015bft,Chen:2017uof} and concluded that only one state, named $\psi(4230)$, remains as a genuine resonance, while the asymmetric structures and other reported states arise from interference between $\psi(4160)$ and $\psi(4415)$ and the continuum background.

Following the identification of $\psi(4230)$, constructing the higher charmonium spectrum became necessary. The Lanzhou group calculated the mass spectrum using an unquenched potential model, employing $\psi(4230)$ as a key scaling point \cite{Wang:2019mhs}. However, the calculated mass of the $\psi(4S)$ state was larger than the experimental value, prompting an investigation into the $4S$–$3D$ mixing scheme. The results suggest the existence of a partner state of $\psi(4230)$, the $\psi(3D)$ named $\psi(4380)$, and indicate a large mixing angle between $\psi(4230)$ and $\psi(4380)$. Furthermore, the $S$–$D$ mixing scheme was revisited in Refs. \cite{Man:2025zfu} using a coupled-channel mechanism, supporting the previous conclusions.

After establishing $\psi(4S)$ and its partner $\psi(3D)$, the nature of the higher $\psi(4415)$ state remains to be clarified. The assignment of $\psi(4415)$ was also studied in Ref. \cite{Wang:2019mhs}. The results suggest that $\psi(4415)$ is a $\psi(5S)$ state, with its corresponding partner $\psi(4D)$ predicted to be $\psi(4500)$; their $S$–$D$ mixing scheme was also discussed to explain their masses. Notably, the BESIII Collaboration indeed discovered a new structure named $Y(4500)$ \cite{BESIII:2022joj}, which was identified as the predicted $\psi(4500)$ \cite{Wang:2022jxj}. In conclusion, with these six vector charmonia mentioned above, the vector charmonium family can be well-established in the $4.0-4.5$ GeV range. This framework provides an explanation for a wide range of experimental phenomena. For instance, the asymmetric structure $Y(4230)$ reported in $e^+ e^- \to \eta J/\psi$ \cite{Peng:2024xui} is described by the constructed charmonium states. Moreover, results from other processes, such as $e^+ e^- \to \pi^+ D^0 D^{*-}$ \cite{Wang:2023zxj} and $e^+ e^- \to \psi(3686) \pi^+ \pi^-$ \cite{Huang:2019agb}, are also consistent with this vector charmonium picture.

There are two high-lying charmonium-like states around 4.6 GeV, the $Y(4660)$ \cite{Belle:2007umv} and the $Y(4630)$ \cite{Belle:2007umv}, they were first reported by the Belle Collaboration. Following the analysis of $\psi(4230)$ and $\psi(4415)$, the Lanzhou group computed the mass spectrum of charmonia above 4.6 GeV \cite{Wang:2020prx} using the same approach as in Ref. \cite{Wang:2019mhs}. Similar to the cases of $Y(4260)$ and $Y(4360)$, the results suggest that the $Y(4660)$ and $Y(4630)$ arise from the interference between two charmonium-like structures, $Y(4585)$ and $Y(4676)$. Furthermore, Ref. \cite{Wang:2020prx} identified these structures as high-lying charmonia within a $6S$–$5D$ mixing scheme and predicted states up to 4.9 GeV. The subsequent observation of the $Y(4710)$ by BESIII \cite{BESIII:2023wqy} is consistent with the prediction.

Among the higher orbital $D$-wave states, $\psi_2(3823)$ \cite{Belle:2013ewt} and $\psi_3(3842)$ \cite{LHCb:2019lnr} have been identified. The mass of $\psi_2(3823)$ lies below the $D\bar{D}^*$ threshold, and its main decay modes are hidden-charm decays. $\psi_3(3842)$, discovered in proton-proton collisions, predominantly decays to $D\bar{D}$. The spectrum of higher $D$-wave states remains incomplete, and $F$-wave charmonia await future discovery.

Currently, the experimental understanding of high-lying charmonium remains relatively limited compared to light flavored mesons. Facilities such as BESIII, Belle II, LHCb, and the planned Super $\tau$–Charm Facility (STCF) will be instrumental in searching for these high-lying charmonium states and expanding our knowledge of these undiscovered resonances. The subsequent section will discuss the most promising candidates for detection among these states. Therefore, it is necessary to extend our study and predict the higher radial and orbital charmonium spectrum using the same model as in Ref. \cite{Wang:2019mhs}.

\section{Theoretical framework}\label{sec3}

\subsection{The MGI model for calculating mass spectrum} \label{MGI}

In this study, we employ the MGI model to calculate the mass spectrum of the 
high-lying $2D$ and $1F$ charmonia. To obtain the mass spectrum, we introduce a screening potential within the MGI framework \cite{Song:2015nia, Song:2015fha, Pang:2017dlw, Wang:2018rjg, Wang:2019mhs, Wang:2020prx, Wang:2021abg}. The relevant Hamiltonian is

\begin{align}\label{2.1}
\tilde{H}=(p^2+m_1^2)^{1/2}+(p^2+m_2^2)^{1/2}+\tilde{V}_{\rm eff}(\boldsymbol{p},\boldsymbol{r}),
\end{align}
where $m_1$ and $m_2$ are equal and represent the mass of the $c$ or $\bar{c}$ quark, respectively. The effective potential $\tilde{V}_{\rm eff}(\boldsymbol{p}, \boldsymbol{r})$ describes  the interaction between $c$ and $\bar{c}$, including both a short-range one-gluon-exchange term $\gamma^{\mu}\otimes\gamma_{\mu}$ and a long-range confinement term $1\otimes1$. In the non-relativistic limit, $\tilde{V}_{\rm eff}(\boldsymbol{p},\boldsymbol{r})$ is reduced to the  nonrelativistic potential $V_{\rm eff}(r)$:

\begin{align}\label{2.2}
V_{\rm eff}(r)=H^{\rm conf}+H^{\rm hyp}+H^{\rm so},
\end{align}
where
\begin{align}\label{2.3}
H^{\rm conf}=br-\frac{4\alpha_s(r)}{3r}+c
\end{align}
is the spin-independent potential, which includes the confining potential, the Coulomb-like potential, and a constant term. Here, $\alpha_s(r)$ is the running coupling constant. The color-hyperfine interaction $ H^{\text {hyp }}$ in Eq. \eqref{2.2} consists of the spin-spin and  tensor terms, given by
\begin{equation}\begin{aligned}\label{2.4}
    H^{\text {hyp }}= & \frac{4\alpha_s(r)}{3m_1 m_2}\left[\frac{8 \pi}{3} \boldsymbol{S}_1 \cdot \boldsymbol{S}_2 \delta^3(\boldsymbol{r})\right. \\
    & \left.+\frac{1}{r^3}\left(\frac{3 (\boldsymbol{S}_1 \cdot \boldsymbol{r})( \boldsymbol{S}_2 \cdot \boldsymbol{r})}{r^2}-\boldsymbol{S}_1 \cdot \boldsymbol{S}_2\right)\right],\end{aligned}\end{equation}    
where $\boldsymbol{S}_{1(2)}$ denotes the spin of the quark (antiquark). The spin-orbit interaction in Eq. \eqref{2.2} is expressed as

\begin{align}\label{2.5}
H^{\rm so}=H^{\rm so(cm)}+H^{\rm so(tp)}.
\end{align}
Here, $H^{\rm so(cm)}$ represents the color-magnetic term, and the $H^{\rm so(tp)}$ is the Thomas precession term, which can be written as
\begin{align}\label{2.6}
H^{\rm so(cm)}=\frac{4\alpha_{s}(r)}{3r^{3}}\left(\frac{ \boldsymbol{S}_{1}}{m^2_{1}}+\frac{\boldsymbol{S}_{2}}{m^2_{2}}+\frac{
\boldsymbol{S}_{1}+\boldsymbol{S}_{2}}{m_{1}m_{2}}\right) \cdot\boldsymbol{L},
\end{align}

\begin{align}\label{2.7}
H^{\rm so(tp)}=-\frac{1}{2 r} \frac{\partial H^{\rm scr}}{\partial r}\left(\frac{ \boldsymbol{S}_{1}}{m_{1}^{2}}+\frac{ \boldsymbol{S}_{2}}{m_{2}^{2}}\right) \cdot\boldsymbol{L},
\end{align}
where $\boldsymbol{L}$ is the relative orbital angular momentum between quark and antiquark. Additionally, the smearing transformation and momentum-dependent factors play a dominant role in the relativistic effects within the MGI model. On the one hand, we apply the smearing to the screened potential $S(r)=\frac{b(1-e^{-\mu r})}{\mu}+c$ and the Coulomb-like potential $G(r)=-\frac{4\alpha_s(r)}{3r}$. For simplicity, we use the general symbol $V(r)$ to represent both $G(r)$ and $S(r)$. The smearing transformation is given by

\begin{align}\label{2.8}
\tilde{V}(r)=\int d^3\boldsymbol{r}'\rho(\boldsymbol{r}-\boldsymbol{r}')V(r'),
\end{align}
where
\begin{align}\label{2.9}
\rho(\boldsymbol{r}-\boldsymbol{r}')=\frac{\sigma^3}{\pi^{3/2}}\mathrm{exp}\left[-\sigma^2(\boldsymbol{r}-\boldsymbol{r}')^2\right]
\end{align}
is the smearing function, with $\sigma$ as the smearing parameter. On the other hand, momentum dependent factors are introduced. For the smeared Coulomb-like and smeared spin-dependent term, the semirelativistic corrections are
\begin{equation}\label{2.9}
\begin{split}
    \tilde{G}(r)\to &\left(1+\frac{p^2}{E_1E_2}\right)^{1/2}\tilde{G}(r)\left(1+\frac{p^2}{E_1E_2}\right)^{1/2},\\
    \tilde{V}_i(r)\to &\left(\frac{m_1m_2}{E_1E_2}\right)^{1/2+\epsilon_i}\tilde{V}_i(r)
    \left(\frac{m_1m_2}{E_1E_2}\right)^{1/2+\epsilon_i}.
\end{split}
\end{equation}
 $E_1$ and $E_2$ represent the energies of the c-quark and $\bar{c}$-quark in charmonium, respectively. The correction factors, $\epsilon_i$, account for various types of hyperfine interactions, including spin-spin and tensor terms, as described in Ref. \cite{Godfrey:1985xj}. The momentum dependent factors have been incorporated into Eqs. \eqref{2.4} - \eqref{2.7}. Furthermore, to account for the unquenched effect, it is common to replace the line potential with the screening potential
\begin{align}
br\to\frac{b(1-e^{-\mu r})}{\mu},
\end{align}
where $\mu$ represents the strength of the screening effect. Therefore, the spin-independent potential in the MGI model becomes 

\begin{align}
H^{\rm scr}=\frac{b(1-e^{-\mu r})}{\mu}-\frac{4\alpha_s(r)}{3r}+c.
\end{align}

In our study, we employed the color-screening confinement potential model, constructed in 1995 \cite{Ding:1995he}. This model demonstrates that the linear potential may be screened at large distances due to the creation of a light quark pair. Its formulation is inspired by earlier work in Lattice QCD \cite{Born:1989iv}. With the growing number of observed hadronic states \cite{ParticleDataGroup:2024cfk}, this model has been applied to calculate hadron spectroscopy, including both light-flavor and heavy-flavor meson systems \cite{Song:2015nia, Song:2015fha, Wang:2019mhs, Wang:2020prx, Bai:2025knk, Li:2009ad}, offering an effective approach to account for the unquenched effects.

Currently, there are two primary methods for addressing the unquenched effect in hadron spectroscopy. One is the color-screening confinement potential model mentioned above, while the other involves coupled-channel analysis, which considers the coupling of bare states to allowed hadronic channels. {In the present work, the modified MGI model with a screening potential is employed as an effective means to incorporate unquenched effects in high-lying states. This treatment provides a consistent first-step description of the spectrum. Building upon this foundation, our future studies will adopt a coupled-channel framework to explicitly incorporate channel-coupling dynamics and to achieve a more complete understanding of the hadronic states discussed above.}


The MGI model parameters adopted in this study are the same as those in Ref. \cite{Wang:2019mhs}, and they effectively reproduce the observed charmonium mass spectrum. These parameters are listed in Table \ref{Parameters}. The mass spectrum and spatial wave functions are determined by solving the Schrödinger equation with the MGI potential and the specified parameters. Here, we focus on $2D$ and $1F$ charmonium states, where an unquenched model is required to account for screening effects.


\begin{table}[!htbp]\centering
    \captionsetup{justification=raggedright, singlelinecheck=false}
    \caption{The parameters of the MGI model used in this work. These parameters are determined from our previous work \cite{Wang:2019mhs}.
    }
    \renewcommand\arraystretch{1.2}
    \label{Parameters}
    \begin{tabular*}{85mm}{c@{\extracolsep{\fill}}lclc}
    \toprule[0.5pt]\toprule[0.5pt]
    \text{Parameters}   &\text{Values}      &\text{Parameters}    &\text{Values}\\
    \toprule[0.5pt]
    $b$                   & $0.2687~\text{GeV}^2$ &$\epsilon_t$&0.012&\\
    $c$                   & $-0.3673$~\text{GeV}    &$\epsilon_{\text{so(V)}}$&$-0.053$&\\
    $\mu$                 &$ 0.15~\text{GeV} $    &$\epsilon_{\text{so(S)}}$&$0.083$&\\
    $\epsilon_c$          & $-0.084$&$m_{u}$&$0.22~\text{GeV}$  &\\
    $m_{d}$          & $0.22~\text{GeV}$ &$m_{c}$&$1.65~\text{GeV}$  &\\
    \bottomrule[0.5pt]\bottomrule[0.5pt]
    \end{tabular*}
    \end{table}

\subsection{The QPC model for the study of OZI-allowed strong decays} \label{QPC}

To calculate the strong decay widths of these discussed charmonium states, the
quark pair creation (QPC) model \cite{Micu:1968mk,LeYaouanc:1973ldf,LeYaouanc:1977gm,LeYaouanc:1977fsz} is utilized in this work. It has been successfully applied to the quantitative description of the OZI-allowed strong decay of hadrons \cite{Wang:2019qyy, Chen:2019ywy, Luo:2023sne, Pang:2017dlw, Li:2022bre, Wang:2019mhs,Liu:2024xav}. In the QPC model, the transition matrix for the $A \rightarrow B+C$ process is written as $\langle B C|\mathcal{T}| A\rangle=\delta^3\left(\boldsymbol{P}_B+\boldsymbol{P}_C\right)$ $\mathcal{M}^{M_{J_A} M_{J_B} M_{J_C}}(\boldsymbol{P})$, where $\mathcal{M}^{M_{J_A} M_{J_B} M_{J_C}}(\boldsymbol{P})$ represents the helicity amplitude, and $\boldsymbol{P}_B$ and $\boldsymbol{P}_C$ are the momenta of mesons $B$ and $C$, respectively, in the stationary reference frame of meson $A$. The states $|A\rangle,|B\rangle$, and $|C\rangle$ refer to the mock states associated with mesons $A, B$, and $C$. The transition operator $\mathcal{T}$ describes the quark-antiquark pair creation from the vacuum, and it has the form

\begin{equation}\label{3.1}
    \begin{aligned}
    \mathcal{T}= & -3 \gamma \sum_{m, i, j}\langle 1 m ; 1-m \mid 00\rangle \int d \boldsymbol{p}_3 d \boldsymbol{p}_4 \delta^3\left(\boldsymbol{p}_3+\boldsymbol{p}_4\right) \\
    & \times \mathcal{Y}_{1 m}\left(\frac{\boldsymbol{p}_3-\boldsymbol{p}_4}{2}\right) \chi_{1,-m}^{34} \phi_0^{34}\left(\omega_0^{34}\right)_{i j} b_{3 i}^{\dagger}\left(\boldsymbol{p}_3\right) b_{4 j}^{\dagger}\left(\boldsymbol{p}_4\right),
    \end{aligned}
    \end{equation}    
where the dimensionless constant $\gamma$ describes the intensity of quark pairs $u\bar{u}$, $d \bar{d}$, or $s \bar{s}$ produced from the vacuum and can be determined from experimental data \footnote{In fact, the parameters of the MGI model are determined by observed charmonium mass spectra, its confinement and static corrections arising from sea quark-antiquark pairs in the QCD vacuum. Crucially, this model does not explicitly include dynamical quark-pair creation or decay amplitudes. On the other hand, the QPC model describes the dynamical process of light quark-antiquark pair creation during two-body strong decay. The quark-pair creation parameter $\gamma$  is determined by fitting experimental partial decay widths of charmonium states, and is independent of the screened potential’s spectroscopic inputs. Two models address separate observables (masses vs. decay widths) and their parameters are derived from entirely independent datasets.} 

The $\chi_{1,-m}^{34}$ is a spin-triplet configuration, while $\phi_0^{34}$ and $\omega_0^{34}$ represent the SU(3) flavor and color singlets, respectively. The term $\mathcal{Y}_{l m}(p)=|p|^l Y_{l m}(p)$ is the solid harmonic function. 

The helicity amplitude ${\cal M}^{M_{J_A}M_{J_B}M_{J_C}}(\boldsymbol{P})$ can be related to the partial wave amplitude using the Jacob-Wick formula \cite{Jacob:1959at}

 \begin{equation}\label{3.2}
    \begin{aligned}
    {\cal M}^{JL}(A\rightarrow BC)& =\frac{\sqrt{4\pi(2L+1)}}{2J_A+1}\sum_{M_{J_B}M_{J_C}}
    \langle L0;JM_{J_A}|J_AM_{J_A}\rangle\\
    & \quad \times \langle J_BM_{J_B};J_CM_{J_C}|J_AM_{J_A}\rangle\\
    & \quad \times {\cal M}^{M_{J_A}M_{J_B}M_{J_C}}(\boldsymbol{P}),
    \end{aligned}
    \end{equation} 
where $\boldsymbol{L}$ is the orbital angular momentum between the final states $B$ and $C$, and $\boldsymbol{J}=\boldsymbol{J}_B+\boldsymbol{J}_C$. The general partial width for the $A \rightarrow BC$ decay is

\begin{equation}\label{3.3}
    \Gamma_{A\rightarrow BC}=\frac{\pi^2}{4}\frac{|\boldsymbol{P}|}{m^2_A}\sum_{J,L}|{\cal M}^{JL}(\boldsymbol{P})|^2,
\end{equation}
where ${m}_{A}$ is the mass of the parent meson $A$. The dimensionless parameter $\gamma=5.84$ is the same as in Ref. \cite{Wang:2019mhs}, and the strength for creating $s\bar{s}$ from the vacuum satisfies the relation of $\gamma_s=\gamma/\sqrt{3}$.  


{In our framework, the uncertainties in the decay widths are primarily determined by the quark pair-creation strength $\gamma$ in the QPC model, which significantly influences the decay widths. We adopted the central value $\gamma = 5.84$ from Ref. \cite{Wang:2019mhs}, which has been successfully applied to high-lying charmonia. To assess the sensitivity, we have now varied $\gamma$ by $\pm 0.5$ around this central value, representing the typical systematic uncertainty associated with the pair-creation strength in this energy region. }

For our calculations, we use the numerical spatial wave functions obtained in Section \ref{MGI} as inputs. The numerical spatial wave functions of the final mesons are the same as those used in  Refs. \cite{Song:2015nia,Song:2015fha}. The results of our calculations are presented in Tables \ref{Dtable} and \ref{Ftable}.

\subsection{Radiative decay} \label{Rad}

In this section, we briefly outline the model used to calculate radiative decay. The quark-photon electromagnetic coupling is described by
\begin{align}\label{4.1}
        H_e=-\sum_j e_j \bar{\psi}_j \gamma_\mu^j A^\mu(\mathbf{k}, \mathbf{r}) \psi_j,
\end{align}
where $\psi_j$ is the $j$-th quark field with a charge $e_j$ in a hadron, and $\mathbf{k}$ denotes the 3-momentum of the photon.

The spatial wave functions are calculated using the potential models outlined in Sec. \ref{MGI}. The nonrelativistic expansion of $H_e$ can be written as \cite{Brodsky:1968ea,Close:1970kt,Li:1994cy,Li:1997gd,Deng:2016stx,Wang:2017kfr}
\begin{equation}\label{4.2}
    h_e \simeq \sum_j\left[e_j \mathbf{r}_j \cdot \boldsymbol{\epsilon}-\frac{e_j}{2 m_j} \boldsymbol{\sigma}_j \cdot(\boldsymbol{\epsilon} \times \hat{\mathbf{k}})\right] e^{-i \mathbf{k} \cdot \mathbf{r}_j},
    \end{equation}
where $\boldsymbol{\sigma}_j$, $ m_j$, and $\mathbf{r}_j$ stand for Pauli spin vector, the constituent mass and the coordinate for the $j$-th quark, respectively. The vector $\boldsymbol{\epsilon}$ is the polarization vector of the photon. The standard helicity transition amplitude $\mathcal{A}_\lambda$ between the initial state $|J_i \lambda_i\rangle$ and the final state $\left|J_f \lambda_f\right\rangle$ is given by
\begin{equation}\label{4.3}
    \mathcal{A}_\lambda=-i \sqrt{\frac{\omega_\gamma}{2}}\left\langle J_f \lambda_f\left|h_e\right| J_i \lambda_i\right\rangle,
    \end{equation}
where $\omega_\gamma$ is the photon energy. $J_f$ and $J_i$ are the total angular momenta of the final and initial mesons, respectively, and $\lambda_f$ and $\lambda_i$ are the components of their total angular momentum along the $z$ axis. In our calculations, we choose the initial hadron-rest frame for the radiative decay process, so that the momentum of the initial hadron is $\mathbf{P}_i=0$, and the final hadron's momentum is $\mathbf{P}_f=-\mathbf{k}$. We set the polarization vector of the photon as $\boldsymbol{\epsilon}=-\frac{1}{\sqrt{2}}(1, i, 0)$, with the photon momentum directed along the $z$ axial $(\mathbf{k}=k \hat{\mathbf{z}})$.  

The partial decay widths for the electromagnetic transitions are given by
\begin{equation}\label{4.4}
    \Gamma=\frac{|\mathbf{k}|^2}{\pi} \frac{2}{2 J_i+1} \frac{M_f}{M_i} \sum_\lambda\left|\mathcal{A}_\lambda\right|^2,
    \end{equation}
where $J_i$ is the total angular momenta of the initial mesons, and $M_i$ and $M_f$ are the masses of the initial and final charmonium states, respectively.
The electromagnetic transition rates of predominant decay modes we calculated are presented in Tables \ref{Dtable} and \ref{Ftable}.

\section{The missing  $2D$ and $1F$-wave charmonium states}\label{sec4}

The masses of the $D$-wave and $F$-wave charmonium states obtained using the MGI models are listed in Table \ref{Dtable}. According to the heavy quark spin symmetry, the charmonium states with the same radial number are approximately degenerate and exhibit similar properties. In our calculation, the masses of $2D$ states are approximately 4140 MeV, below the mass of $2^3D_1$ state is 4125 MeV, lower than the value listed in the PDG \cite{ParticleDataGroup:2024cfk}. As discussed in Sec. \ref{sec2}, a lower mass for the $\psi(4160)$ is supported by recent analyses \cite{Peng:2024blp, Man:2025vmm}, and our result is consistent with these findings. The near-degeneracy of the other $2D$ state masses is consistent with expectations from heavy quark symmetry.

The calculated masses of the $1F$ states exhibit near-degeneracy at approximately 4070 MeV, reflecting the heavy quark spin symmetry in the $F$-wave sector. However, the mass of the $1^3F_2$ state is slightly lower than those of the other $F$-wave states by about 5 MeV. Furthermore, their proximity to the $D^*\bar{D}^*$ threshold suggests a potential enhancement of the decay width in this channel.

We observe distinct mass splittings within certain charmonium multiplets: the $2^3D_1$ mass is lower than the other 2D states, and the $1^3F_2$ mass is lower than its near-degenerate partners. This pattern is consistent with the established behavior of the $1^3P_0$ state, whose mass is lower than other $1P$ states according to experimental data \cite{ParticleDataGroup:2024cfk}. These systematic deviations arise from the spin-orbit interaction. According to Eq. \eqref{2.4}, the spin-orbit terms are substantial and negative for these states, leading to a reduction in their predicted masses. In such cases, the inclusion of state mixing can increase the final masses. For instance, the $4S$–$3D$ mixing scheme studied in Ref.~\cite{Wang:2019mhs} raised the mass of the $3D$ state from about 4334 MeV to 4380 MeV.

Heavy quark spin symmetry is more manifest in states with higher radial quantum numbers. This is evidenced by the $1F$ states exhibiting greater degeneracy than the $1D$ states, and the $2D$ states being more degenerate than the $2P$ states. This phenomenon is attributable to two factors. Firstly, the symmetry is more prominent at higher masses, which are associated with larger radial quantum numbers. Secondly, a larger radial number corresponds to a greater mean radius $r$. Since the spin-orbit and color-hyperfine interactions (Eqs. \eqref{2.4} - \eqref{2.7}) are inversely correlated with $r$, their effects are suppressed in these states, reducing symmetry-breaking contributions.

\begin{table}
\centering
\captionsetup{justification=raggedright, singlelinecheck=false}
\caption{The open charm decay behaviors and major radiative decay channels of the $2D$ states. When the two-body strong decay channels are not open or forbidden, a symbol ``\ding{55}" is presented. {The superscript and subscript of the decay widths represents the upper and lower limits of the decay widths corresponding to $\gamma = 5.84 \pm 0.5$.} }\label{Dtable}
\begin{adjustbox}{center}
{\fontsize{7.2}{11}\selectfont
\begin{tabular}{cc cc cc cc}
\toprule \toprule
&  &  \multicolumn{2}{c}{   $2^1D_2$  }      &  \multicolumn{2}{c}{   $2^3D_2$  }    &  \multicolumn{2}{c}{   $2^3D_3$  }      \\              
 \multicolumn{2}{c}{Mass } &   \multicolumn{2}{c}{  4137  }  &  \multicolumn{2}{c}{  4137  }   &  \multicolumn{2}{c}{  4144  }  \\   \hline                    
\multicolumn{8}{c}{Strong  decay}\\
\hline  
 
&    Mode     & $ \Gamma_{thy}$\ (\text{MeV}) &  Br  &   $ \Gamma_{thy}$\ (\text{MeV}) &  Br  &   $ \Gamma_{thy}$\ (\text{MeV})  &  Br  \\ 

\multicolumn{2}{c}{$D\bar{D}$}  & ~\ding{55}  &   & ~\ding{55}  &   &$3.3^{+0.7 }_{-0.5}$&  6.0\%  \\
 
\multicolumn{2}{c}{$D \bar{D}^*$\footnotemark[1]}  & $29.5^{+5.0}_{-5.0}$ & $51.0$\%  &$22.9^{+3.8}_{-4.0}$ & $47.1 $\%  &$21.8 ^{+3.8}_{-3.6}$&  $40.1$\%  \\
 
\multicolumn{2}{c}{$D^*\bar{D}^*$}   & $27.2^{+4.6}_{-4.7}$ & 47.0\%  &$24.0^{+4.8}_{-3.6} $ &  $49.3$\%  &$28.8^{+3.8}_{-5.7}$&  $53.0$\%\\
 
\multicolumn{2}{c}{$D_s \bar{D}_s$}   & ~\ding{55}  &     & ~\ding{55}  &     &$0.4^{+0..02}_{-0.1}$& $ 0.7 $\% \\
 
\multicolumn{2}{c}{$D_s \bar{D}_s^*$}   & $1.2^{+0.2}_{-0.2}$ & $2.0 $\% &$1.8^{+0.4}_{-0.3}$& $3.6 $\% &$0.1^{+0.01}_{-0.03}$&  $0.2 $\% \\

\multicolumn{2}{c}{ Total} &$57.9^{+9.8}_{-9.9}$ & 100\%  &$48.7^{+9.0}_{-7.9}$&  100\%   &$54.4^{+8.3}_{-9.9}$&  100\% \\
\hline   
\multicolumn{8}{c}{Radiative  decay}\\
\hline   
 && Mode & $ \Gamma_{thy}~(\text{keV})$ & Mode & $ \Gamma_{thy}~(\text{keV})$ & Mode & $ \Gamma_{thy}\ (\text{keV})$ \\
 & & $2^1P_1\gamma$ & 167 &  $2^3P_1\gamma$  & 128 & $2^3P_2\gamma$ & 213 \\
&      &  $1^1P_1\gamma$ & 58.45 & $2^3P_2\gamma$ & 63 & $1^3P_2\gamma$ & 41 \\
\bottomrule  \bottomrule 
\end{tabular}
}\end{adjustbox}
\footnotetext[1]{In this work, the notation $D\bar{D}^*$ is used as a shorthand that encompasses both charge-conjugate pairs: $D\bar{D}^*$ and $\bar{D}D^*$. All other open-charm meson pairs are denoted analogously.} 
\end{table}

\subsection{The missing $D$-wave states} \label{Dwave}

This work does not revisit the $2^3D_1$ charmonium state and the $S$–$D$ mixing scheme, which has been extensively analyzed. More details of our previous study of $2^3D_1$ state and the $3S$–$2D$ mixing scheme can be found in Ref. \cite{Peng:2024blp,Man:2024mvl, Man:2025vmm}. The present focus is the missing $2^1D_2$, $2^3D_2$, and $2^3D_3$ states. We systematically analyze their OZI-allowed strong decay and radiative decay with the results presented in Table \ref{Dtable}.

\begin{figure*}[htbp]
	\centering
	\begin{tabular}{c}
		\includegraphics[width=1\textwidth]{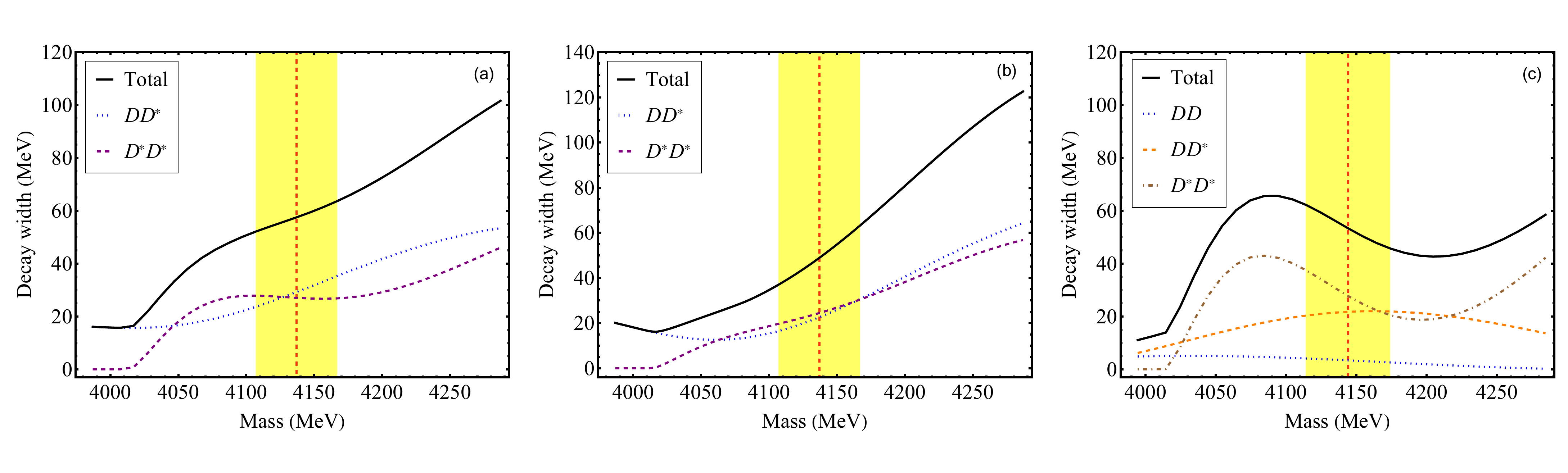}
	\end{tabular}
    \captionsetup{justification=raggedright, singlelinecheck=false}
	\caption{ Strong decay widths of the $D$-wave charmonium states near their open-flavor thresholds. Figures (a), (b), and (c) display the partial widths as a function of mass for the $\eta_{c2}(4137)$, $\psi_2(4137)$, and $\psi_3(4144)$ states, respectively. The horizontal axis spans $\pm 150$ MeV around the nominal mass of each state. Decay channels with partial widths below 5 MeV are omitted for clarity but are included in the total width sums. The yellow band marks the $\pm 30$ MeV region around our calculated central mass values, and the red dashed line indicates the precise calculated mass. }
	\label{Fig3}
\end{figure*}

The masses of $2D$ and $1F$ states are close to the $D^*\bar{D}^*$ threshold, therefore we present the decay behaviors of these states near open flavor threshold in Figs. \ref{Fig3}-\ref{Fig4}. The mass axis spans $\pm150$ MeV around the central value. Compared to previous results of Refs. \cite{Deng:2016stx, Li:2009zu} from other unquenched models, the discrepancy between our results and other values does not exceed 30 MeV. Consequently, Consequently, the range of $\pm30$ MeV around the calculated mass is adopted as an estimate for the investigated states.

\paragraph*{The $\eta_{c2} (4137)$.}In our calculation, the masses of the $2D$ charmonium states are around 4140 MeV. Their open-flavor decay channels as functions of mass are shown in Fig. \ref{Fig3}. For the $2^1D_2$ state, with a mass of 4137 MeV, is denoted as $\eta_{c2}(4137)$. According to Table \ref{Dtable}, its dominant decay modes are $D\bar{D}^*$ and $D^*\bar{D}^*$, which together account for about 98\% of the total branching fraction, and specifically, 51\% for $D\bar{D}^*$ and 47\% for $D^*\bar{D}^*$. We therefore suggest that the $2^1D_2$ charmonium state could be identified experimentally through these channels. The $D_s \bar{D}_s^*$ channel also contributes, though with a relatively small branching ratio of about 2\%. The total decay width of $\eta_{c2}(4137)$ is approximately 58 MeV, which is slightly smaller than the measured width of $\psi(4160)$. 

The radiative decay widths of the $D$-wave states, listed in Table \ref{Dtable}, show that transitions to $P$-wave and $F$-wave final states are generally stronger, while those to $S$-wave and $D$-wave states are relatively weaker. For the $2^1D_2$ state, the dominant radiative decay channels are $2^1D_2 \to 2^1P_1\gamma$ and $2^1D_2 \to 1^1P_1\gamma$, with partial widths of 167 keV and 59 keV, respectively. The branching ratios reach 0.3\% and 0.1\%, respectively.


The $X(4160)$ was first observed by the Belle Collaboration in $e^+e^- \to J/\psi D^{(*)} \bar{D}^{(*)}$ process \cite{Belle:2007woe}, and later LHCb Collaboration also reported the $X(4160)$ in the $B^+ \to J/\psi \phi K^+$ process with a significance of more than $4 \sigma$ \cite{LHCb:2021uow}. Its preferred quantum numbers are $2^{-+}$, with a measured mass and width of $4146 \pm 18 \pm 33$ MeV and $135 \pm 28^{+59}_{-30}$ MeV, respectively. If $X(4160)$ is indeed the $2^1D_2$ charmonium state, our calculated mass of 4137 MeV is consistent with experiment, though the predicted total width of 58 MeV is somewhat smaller.

We anticipate that future experiments will provide more precise data to further clarify the properties of the $\eta_{c2}(2D)$ state. According to our analysis, the most promising discovery channels are $D\bar{D}^*$ and $D^*\bar{D}^*$, or radiative transitions such as $2^1D_2 \to 2^1P_1\gamma$ and $1^1P_1\gamma$. 

\paragraph*{The $\psi_{2} (4137)$.} For the $2^3D_2$ state, we obtain a mass of 4137 MeV, identical to that of the $\eta_{c2}(4137)$. As the first excited state of $\psi_2(3823)$, we designate it as $\psi_2(4137)$. Its dominant decay channels are $D\bar{D}^*$ and $D^*\bar{D}^*$, with partial widths of 22.9 MeV and 24.0 MeV, corresponding to branching ratios of 47\% and 49\%, respectively. The total decay width is 48.7 MeV, which is smaller than those of the other $2D$ states. However, Fig. \ref{Fig3} shows a strong mass dependence: varying the mass within the $\pm 30$ MeV  band (marked in yellow, approximately $4100–4170$ MeV) would approximately double the total width. The $D_s\bar{D}_s^*$ channel contributes a width of 1.8 MeV (Br = 3.7\%), which is relatively small. We note that the decay patterns of the $2^1D_2$ and $2^3D_2$ states are very similar, likely due to their nearly degenerate masses. According to Table~\ref{Dtable}, the dominant radiative decay is $2^3D_2 \rightarrow 2^3P_1 \gamma$, with a width of 128 keV and a branching ratio of 0.26\%.


\paragraph*{The $\psi_{3} (4144)$.}
The calculated mass of the $2^3D_3$ state is 4144 MeV. We therefore denote it as $\psi_3(4144)$, identifying it as the first excitation of $\psi_3(3842)$. Its decay is dominated by the $D\bar{D}^*$ and $D^*\bar{D}^*$ channels, with branching ratios of 40\% and 53\%, respectively. The $D\bar{D}$ channel is notably smaller, differing by about an order of magnitude. Decay modes involving strange quarks remain minor, consistent with other $2D$ states. The total width is 54.4 MeV. Although the calculated masses of the $2D$ states differ slightly, their decay patterns and branching ratios are largely similar. Fig. \ref{Fig3} demonstrates that the total width of the $2^3D_3$ state is primarily governed by the $D^*\bar{D}^*$ channel across the considered mass range, while the $D\bar{D}^*$ and $D\bar{D}$ contributions show little mass dependence. Interestingly, the total width decreases with increasing mass in our calculation, a behavior that may be attributed to a node effect in the wave function \cite{Duan:2020tsx}. The dominant radiative decay is $2^3D_3 \to 2^3P_2\gamma$, with a width of 213 keV and a branching ratio of 0.4\%; all other radiative channels are significantly smaller.


For the $2^1D_2$, $2^3D_2$, and $2^3D_3$ states, the dominant hadronic decays are $D \bar{D}^*$ and $D^*\bar{D}^*$, which collectively account for a large fraction of the total width. In comparison, decay modes involving strange mesons are less prominent in our results, underscoring the need for experimental confirmation. As shown in Fig.~\ref{Fig3}, the decay widths of these $2D$ states exhibit a strong dependence on their masses, highlighting the importance of more precise mass measurements in future studies. 

In addition, although not dominant, the radiative decay of the $2D$ states still cannot be ignored. Since the $2^3D_2$, $2^1D_2$, and $2^3D_3$ states have not yet been observed, we suggest further investigation into the radiative decay channels involving the $2P$ states. The dominant radiative decay processes exhibit branching ratios on the order of 0.1\%. These decay modes represent viable and promising avenues for the future observation of the $2D$ states as experimental precision and statistics continue to improve. 

\subsection{The $F$-wave states} \label{Fwave}

Our study of the $1F$ charmonium states comprises the $1^1F_3$, $1^3F_2$, $1^3F_3$, and $1^3F_4$, and extends the same methodology applied to the $2D$ states, covering OZI-allowed strong decays and radiative decays. The corresponding results are listed in Tables~\ref{Ftable}. 

\begin{figure*}[htbp]
	\centering
	\begin{tabular}{c}
		\includegraphics[width=0.66\textwidth]{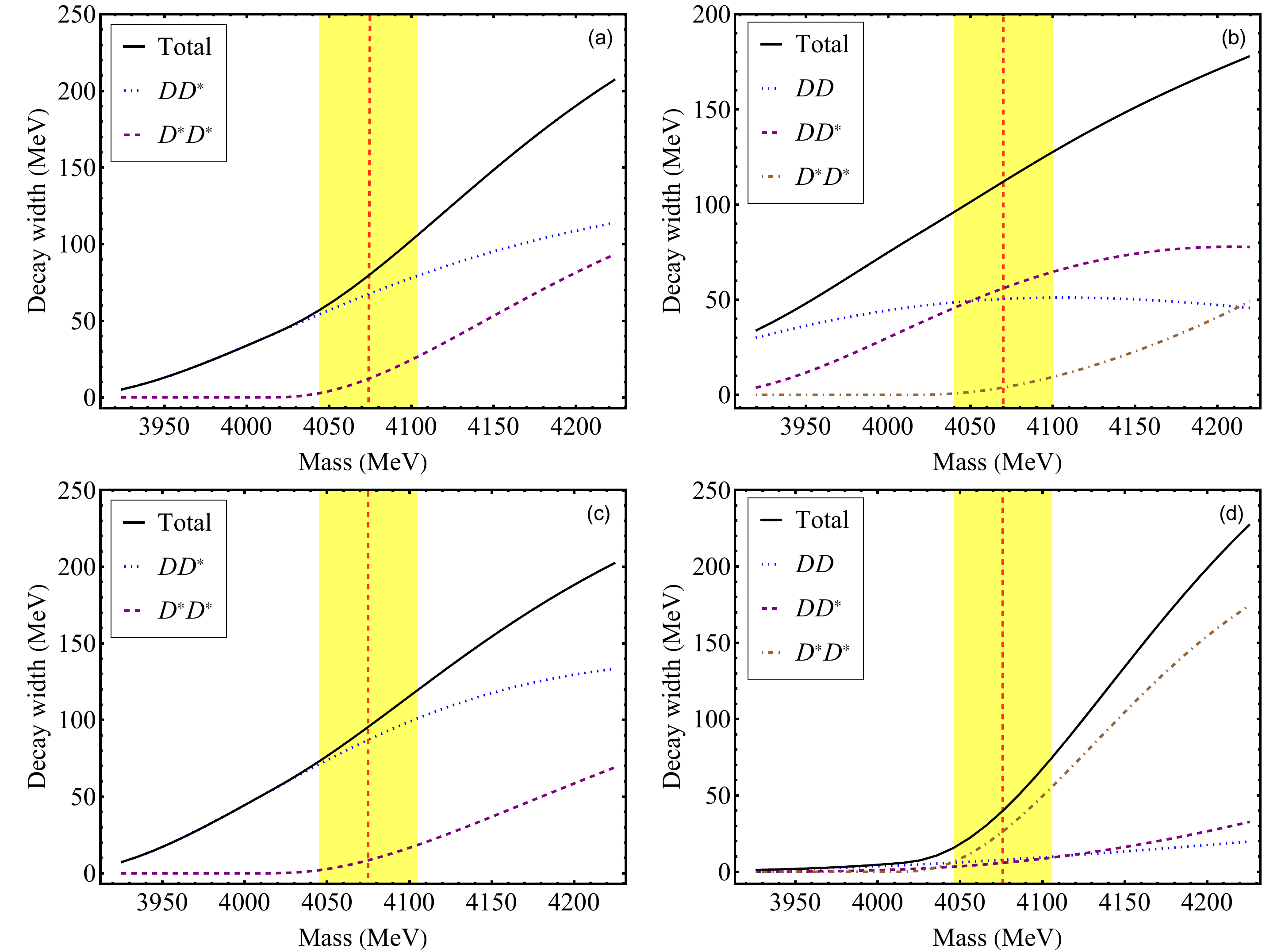}
	\end{tabular}
    \captionsetup{justification=raggedright, singlelinecheck=false}
	\caption{ Strong decay widths of the $F$-wave charmonium states near their open-flavor thresholds. Figures (a), (b), (c), and (d) display the partial widths as a function of mass for the $h_{c3}(4074)$, $\chi_{c2}(4070)$, $\chi_{c3}(4075)$ and $\chi_{c4}(4076)$ states, respectively. The horizontal axis spans $\pm 150$ MeV around the nominal mass of each state. Decay channels with partial widths below 5 MeV are omitted for clarity but are included in the total width sums. The yellow band marks the $\pm 30$ MeV region around our calculated central mass values, and the red dashed line indicates the precise calculated mass. }
	\label{Fig4}
\end{figure*}

The masses of $1F$ states are approximately 4070 MeV, which lie close to the $D^*\bar{D}^*$ threshold. Consequently, their decay properties exhibit a strong dependence on the precise mass values. As shown in Fig. \ref{Fig4}, the decay widths increase rapidly once the masses of the $1F$ states exceed this threshold. This is evident for the $1^3F_4$ state, whose width roughly triples as its mass increases from 4050 MeV to 4100 MeV. Therefore, precise experimental determination of their masses is crucial.

\begin{table*}
\centering
\captionsetup{justification=raggedright, singlelinecheck=false}
\caption{The open charm decay behaviors and major radiative decay channels of the $1F$ states.  When the two-body strong decay channels are not open or forbidden, a symbol ``~\ding{55}" is presented. {The superscript and subscript of the decay widths represents the upper and lower limits of the decay widths corresponding to $\gamma = 5.84 \pm 0.5$.}}\label{Ftable}
\begin{adjustbox}{center}
{\fontsize{7.2}{11}\selectfont
\begin{tabular}{cc cc cc cc cc}
\toprule \toprule
& &  \multicolumn{2}{c}{   $1^1F_3$  }      &  \multicolumn{2}{c}{   $1^3F_2$  }    &  \multicolumn{2}{c}{   $1^3F_3$  }   &  \multicolumn{2}{c}{   $1^3F_4$  }    \\                 
 \multicolumn{2}{c}{Mass } &   \multicolumn{2}{c}{  4074  }  &  \multicolumn{2}{c}{  4070  }   &  \multicolumn{2}{c}{  4075  } &  \multicolumn{2}{c}{  4076  } \\    
\hline                    
\multicolumn{10}{c}{Strong  decay}\\
\hline  
\multicolumn{2}{c}{Mode}  & $ \Gamma_{thy}$\ (\text{MeV}) &  Br  &   $ \Gamma_{thy}$\ (\text{MeV}) &  Br  &  $ \Gamma_{thy}$\ (\text{MeV}) & Br& $ \Gamma_{thy}$\ (\text{MeV}) &  Br\\ 
\multicolumn{2}{c}{$D\bar{D}$}  & ~\ding{55}  &   & $50.5^{+8.9}_{-8.4}$  &  44.9\% & ~\ding{55} &  & $8.1^{+1.1}_{-1.6}$ & 21.3\%  \\
\multicolumn{2}{c}{$D \bar{D}^*$}  &$67.8^{+11.5}_{-11.6}$  & 86.1\%  & $56.4^{+9.5}_{-9.6}$ &  50.1 \%  & $87.6^{+15.0}_{-14.9}$  &  92.0\% & $5.9^{+0.9}_{-1.1}$ & 15.5\%\\
\multicolumn{2}{c}{$D^*\bar{D}^*$}   & $11.0^{+3.2}_{-0.9}$ & 13.9\%  & $3.4^{+1.1}_{-0.2}$ &  3.1\%  & $7.6^{+2.3}_{-0.6}$ &  8.00\% & $23.9^{+7.0}_{-1.9}$ & 62.9\% \\ 
\multicolumn{2}{c}{$D_s \bar{D}_s$}   & ~\ding{55}  &   & $2.2^{+0.1}_{-0.7}$  &  1.9\%  & ~\ding{55}  &  & $0.1 ^{+0.07}_{-0.08}$ &  0.3\%\\
\multicolumn{2}{c}{ total} & $78.8^{+14.7}_{-12.5}$ & 100\%  & $112.5^{+19.6}_{-19.0}$ &  100\%   & $95.2^{+17.3}_{-15.4}$ &  100\%  &  $38.0^{+9.0}_{-4.6}$ & 100\% \\
\hline                    
\multicolumn{10}{c}{Radiative decay}\\
\hline   
\multicolumn{2}{c}{ }  & Mode & $ \Gamma_{thy}\ (\text{keV})$ & Mode & $ \Gamma_{thy}\ (\text{keV})$ & Mode & $ \Gamma_{thy}\ (\text{keV})$ & Mode & $ \Gamma_{thy}\ (\text{keV})$  \\
\multicolumn{2}{c}{ }  & $1^1D_2\gamma$ & 249 &  $1^3D_1\gamma$  & 440 & $1^3D_2\gamma$ & 295 & $1^3D_3\gamma$ & 269 \\
\bottomrule  \bottomrule 
\end{tabular}
}\end{adjustbox}
\end{table*}

\paragraph*{The $h_{c3} (4074)$.}

 For the $1^1F_3$ state, its mass is 4074 MeV, then it is denoted as $h_{c3} (4074)$ by us. Its dominant strong decay channel is $D\bar{D}^*$, with a partial width of 67.8 MeV and a branching ratio of 86.1\%, followed by the $D^*\bar{D}^*$ channel which has a width of about 11 MeV. Although the mass of $h_{c3}(4074)$ is close to the $D^*\bar{D}^*$ threshold, the $D^*\bar{D}^*$ width is smaller than that of $D\bar{D}^*$. However, as shown in Fig. \ref{Fig4}, the branching ratio of $D^*\bar{D}^*$ channel rises to over 30\% if the mass increases to 4100 MeV. The total width of $1^1F_3$ is 78.8 MeV, and these two channels contribute all decay width. This suggests that $h_{c3}(4074)$ has a high potential for observation in $D\bar{D}^*$ and $D^*\bar{D}^*$ final states. The dominant radiative decay channel of  $h_{c3}(4074)$ is $h_{c3}(4074) \to \gamma 1^1D_2$, with a partial width of about 250 keV and a branching ratio of about 0.3\%. Notably, the final state in this process is the missing $\eta_c (1D)$,  suggesting that the observation of $h_{c3}(4074)$ could provide a pathway to its discovery. 

\paragraph*{The $\chi_{c2} (4070)$.}

The predicted mass of the $1^3F_2$ state is 4070 MeV, and we designate it as $\chi_{c2}(4070)$. Although it shares the same $J^{PC}$ quantum numbers with the $\chi_{c2}(2P)$ state, we do not investigate the $P$–$F$ mixing scheme in this work, as the $1^3F_2$ state has not yet been observed. We anticipate that future experimental advances will provide the data needed to study its properties in detail. The main decay channels of $\chi_{c2}(4070)$ are $D\bar{D}$ and $D\bar{D}^*$, with partial widths of 50.5 MeV and 56.4 MeV, respectively. Among these $1F$ states, the broadest is the $1^3F_2$ state, with a width of 112.5 MeV, corresponding to branching ratios of 45\% for $D\bar{D}$ and 50\% for $D\bar{D}^*$. The $D^*\bar{D}^*$ and $D_s \bar{D}_s$ channels are less prominent, with branching ratios of 3\% and 2\%, respectively. As presented in Fig. \ref{Fig4}, the mass dependence of $\chi_{c2}(4070)$ is less pronounced than that of other $F$-wave states: its total width varies by less than 50 MeV over the mass range of 4050–4100 MeV. This is because the $\chi_{c2}(4070) \to D^*\bar{D}^*$ channel exhibits a relatively weak mass dependence. The width of $\chi_{c2} (4070) \to \gamma 1^3D_1$ channel has a width of 440 keV, which is the largest among all radiative transitions of the $2D$ and $1F$ states, corresponding to a branching ratio of approximately 0.4\%. This is a considerable ratio that could be detectable in future experiments. 

\paragraph*{The $\chi_{c3} (4075)$.}

The  $1^3F_3$ state is denoted as $\chi_{c3} (4075)$ for its mass of 4075 MeV. Its proximity in mass to $h_{c3}(4074)$ leads to similar decay patterns: the $D\bar{D}^*$ channel dominates, accounting for 92\% of the total width, while the $D^*\bar{D}^*$ channel is relatively suppressed with a branching ratio of only 8\%. The total width of $\chi_{c3} (4075)$ is about 95.2 MeV, close to the width of $h_{c3} (4074)$. From Fig. \ref{Fig4}, when the mass of $1^3F_2$ state varies from 4050 MeV to 4100 MeV, the partial width of $D^*\bar{D}^*$ channel remains subdominant compared to other two primary channels. It represent the $D^*\bar{D}^*$ channel contributes fewer decay width to $\chi_{c3} (4075)$ state relative to other $1F$ states. The primary radiative decay channels is the $\chi_{c3} (4070) \to \gamma 1^3D_2$, with a width of 295 keV, and a relatively large branching ratio of about 0.3\%. Similar to the $\chi_{c2} (4070)$, the $\chi_{c3} (4075)$ is highly possible to be observed in radiative decay processes in future experiments.

\paragraph*{The $\chi_{c4} (4076)$.}

The masses of $1F$ states are approximately degenerate, especially for the $1^1F_3$, $1^3F_3$, and $1^3F_4$ states, whose masses are nearly identical.We assign the $1^3F_4$ state at 4076 MeV as $\chi_{c4}(4076)$. This state is relatively narrow, with a total strong decay width of 38.0 MeV, which can be attributed to its high orbital angular momentum. The $1^3F_4$ state predominantly decays into the $D\bar{D}$, $D\bar{D}^*$, and $D^*\bar{D}^*$ channels, with a partial width of 8.1 MeV, 5.9 MeV, and 23.9 MeV corresponding to branching ratios of 21\%, 16\%, and 63\%, respectively. The small phase spaces in the $D^*\bar{D}^*$ decay mode lead to narrow partial widths as the mass of $\chi_{c4} (4076)$ is close to the $D^*\bar{D}^*$ threshold. Consequently, even a slight shift in mass could significantly alter the $D^*\bar{D}^*$ decay width. Fig. \ref{Fig4} reveals that the width of the $1^3F_4$ state depends on the $D^*\bar{D}^*$ channel, when the $D^*\bar{D}^*$ channel opens, the width obviously increases with mass. For example, the width of $1^3F_4$ state is from 11.6 MeV at $M = 4055$ MeV to 66.6 MeV at $M = 4155$ MeV. This sharp threshold dependence highlights the necessity of precisely determining the mass of $1^3F_4$ state.

The major radiative decay channel of $\chi_{c4} (4076)$ is $\chi_{c4} (4076) \to 1^3D_3 \gamma$, with the width of 269 keV and a relatively large branching ratio of about 0.7\%. Regarding the $F$-wave states, we observe from Tables \ref{Ftable} that their primary electromagnetic decay channel is to $D$-wave states. For the $1F$ states, the decay channels with the largest widths follow a trend similar to that of the $D$-wave states. For instance, the most significant decay channel for the $1^3F_2$ state is $1^3F_2 \rightarrow 1^3D_1 \gamma$, with a decay width of 440 keV, which is notably large. For the $1^1F_3$, $1^3F_3$, and $1^3F_4$ states, the major decay channels are to $1^1D_2 \gamma$, $1^3D_2 \gamma$, and $1^3D_3 \gamma$, with decay widths of 249 keV, 295 keV, and 269 keV, respectively. Our results reveal that the broadest partial radiative widths correspond to transitions where both the initial and final states have the same quantum numbers $n$ and $S$ for $D$ and $F$ states, and their total angular momenta differ by one ($\Delta J=1$). Following the gradual discovery of the ground states of $D$-wave charmonium mesons, there is increasing optimism that the $2D$ and $1F$ states will be identified in future experimental efforts.

\section{THE PRODUCTION OF HIGH-LYING CHARMONIUM STATES}\label{sec5}

\begin{figure*}[!htbp]
	\centering
	\begin{tabular}{c}
		\includegraphics[width=1\textwidth]{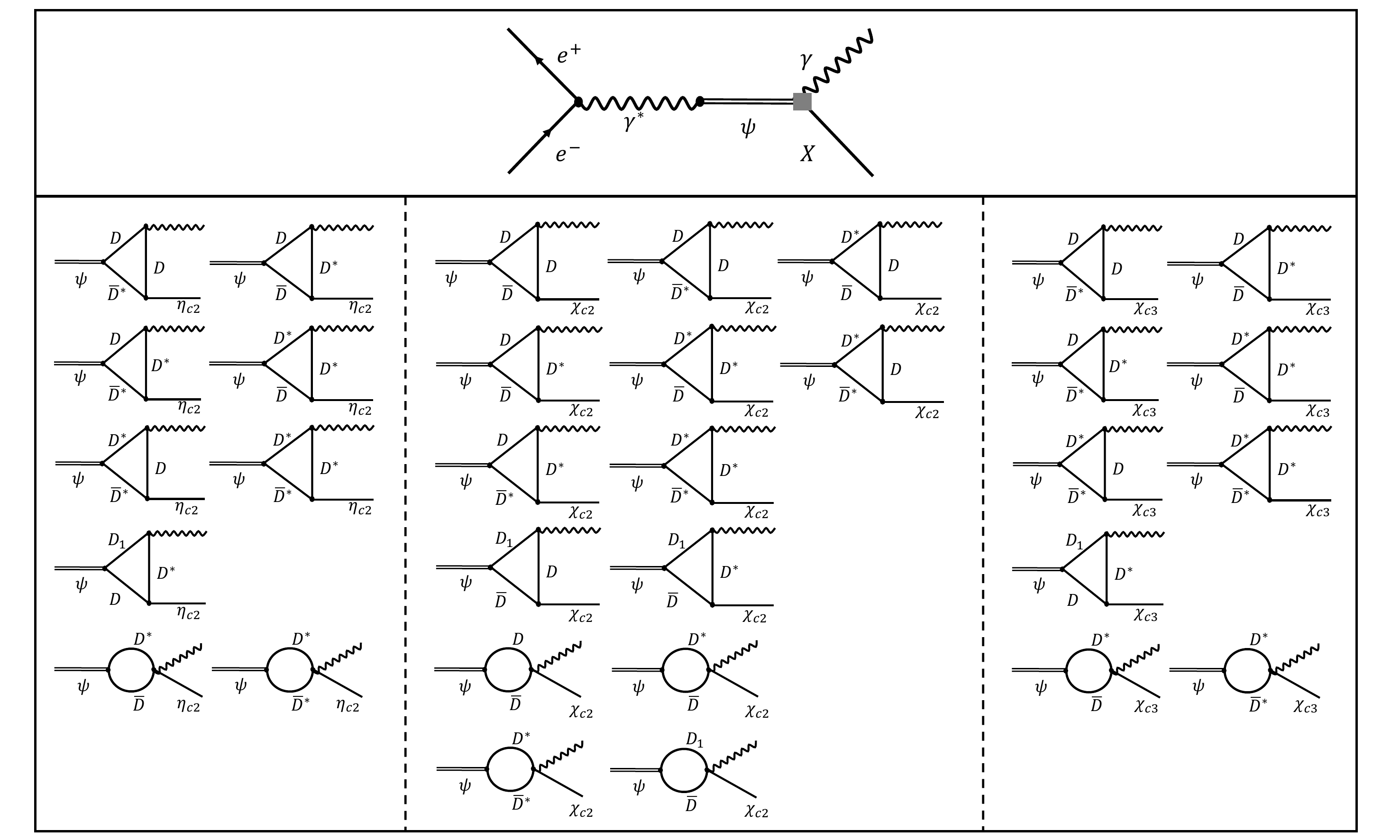}
	\end{tabular}
    \captionsetup{justification=raggedright, singlelinecheck=false}
	\caption{The allowed diagrams of the $e^+e^- \to \psi(4230) \to \gamma X$, including the $\psi(4230) \to \gamma \eta_{c2}(2D)$, $\psi(4230) \to \gamma \chi_{c2}(1F)$, and $\psi(4230) \to \gamma \chi_{c3}(1F)$ processes in the frame of hadronic loop mechanism.}
	\label{Figpro}
\end{figure*}

The $e^+e^-$ annihilation process is a primary mechanism for producing charmonium states, offering a clean environment for discovering $D$-wave and $F$-wave excitations. For example,  the $\psi_2(3823)$ is first observed in $e^+e^- \to \pi^+\pi^-\chi_{c1}\gamma$ \cite{BESIII:2015iqd}. Among the $2D$ states, the $\psi(4160)$ is predominantly observed in $e^+e^-$ collisions \cite{ParticleDataGroup:2024cfk}, and the $X(4160)$—a candidate for the $\eta_c(2D)$ state—was first discovered by Belle in $e^+e^- \to J/\psi D^* \bar{D}^*$ \cite{Belle:2007woe}. Looking ahead, upgrades to major facilities like the Beijing Electron Positron Collider (BEPC) are expected to yield abundant data for $e^+e^-$ processes above 4 GeV. These future datasets will provide excellent opportunities to search for missing $D$- and $F$-wave charmonia. Therefore, in this work, we focus on $e^+e^-$ annihilation processes to evaluate the potential for discovering the $2D$ and $1F$ charmonium states.

The search for the high-lying charmonium states predicted in this work requires center-of-mass energies ($\sqrt{s}$) exceeding their production thresholds, typically above 4.2 GeV. The BESIII experiment has accumulated substantial data in this critical region. A pronounced concentration of integrated luminosity exists around $\sqrt{s} \sim 4.2$ GeV, including dedicated runs at 4.23 GeV (1092 pb$^{-1}$) \cite{BESIII:2013ouc}, 4.26 GeV (826 pb$^{-1}$) \cite{BESIII:2013mhi}, and 4.36 GeV (540 pb$^{-1}$) \cite{BESIII:2013ouc,  BESIII:2017tqk}, complemented by extensive energy scans between 3.81–4.42 GeV \cite{BESIII:2013ouc, BESIII:2017tqk} and 4.19–4.27 GeV \cite{BESIII:2019gjc}. This makes the $\psi(4230)$ resonance, abundantly produced in this region, an ideal initial state for our study. We therefore focus on radiative production via $e^+e^- \to \psi(4230) \to X\gamma$, where $X = \eta_{c2}(2D),\ \chi_{c2}(1F),\ \text{or}\ \chi_{c3}(1F)$. An overview of these pathways is provided in Table \ref{TCP}.

Data at higher energies ($\sqrt{s} \gtrsim 4.4$ GeV), though available, are less abundant. BESIII has performed scans and point-by-point measurements in regions such as 4.29–4.44 GeV \cite{BESIII:2020eyu}, 4.42-4.6 GeV \cite{BESIII:2017tqk, BESIII:2018iea}, 4.61–4.70 GeV \cite{BESIII:2022ulv}, and 4.74–4.95 GeV \cite{BESIII:2022ulv}, with integrated luminosities on the order of several fb$^{-1}$. While crucial for exploring higher-mass states, the current statistics in these regions are not yet sufficient for high-precision studies of the specific channels considered here. Consequently, our quantitative analysis concentrates on the well-surveyed $\sqrt{s} \sim 4.2$ GeV region.

\begin{table*}[htbp]
\centering
\caption{Production processes of $2D$ and $1F$ states in $e^+e^-$ annihilation.}
\label{TCP}
\begin{tabular}{ccccccccccccc}
\toprule \toprule
\multirow{2}{*}{\textbf{states}} & \multirow{2}{*}{\textbf{mode}} & \multirow{2}{*}{\textbf{process}} & \multicolumn{7}{c}{\textbf{$J^{PC}$}} & \multirow{2}{*}{\boldmath $m_\psi$ (GeV)} &  & \\
\cmidrule(lr){4-10}
 & & & $2^{--}$ & $3^{--}$ & $2^{-+}$ & $2^{++}$& $3^{++}$ & $4^{++}$ & $3^{+-}$ &\\
\midrule
\multirow{4}{*}{$2D$} &  radiative decay & $\gamma\eta_c(2^1D_2)$ & $-$ & $-$ & $P$ &$-$&$-$&$-$&$-$& $>4.2$ \\
\cmidrule(lr){2-10}
 & \multirow{3}{*}{hidden-charm decay} & $\pi^+\pi^- \psi(2^3D_J)$ & $D$ & $D$ & $-$&$-$ & $-$&$-$ &$-$ & $>4.66$ \\
 & & $\eta\psi(2^3D_J)$ & $P$ & $F$ &$-$&$-$& $-$&$-$ &$-$ &$>4.71$ \\
 & & $\omega\eta_c(2^1D_2)$ & - & - & $P$ &$-$&$-$&$-$ &$-$ &$>4.94$ \\
\midrule
\multirow{3}{*}{$1F$} &  radiative decay & $\gamma \chi_c(1^3F_J)$ & $-$ & $-$ & $-$ & $S$& $D$ & $D$ &$-$& $>4.1$ \\
\cmidrule(lr){2-10}
 & \multirow{2}{*}{hidden-charm decay} & $\omega\chi_c(1^3F_J)$ & $-$ & $-$ & $-$ & $S$ & $D$ & $D$ & $-$& $>4.85$ \\
 & & $\eta\chi_c(1^1F_J)$ & $-$ & $-$ & $-$ & $D$ & $-$ & $-$& $D$ & $>4.62$ \\
\bottomrule \bottomrule
\end{tabular}
\end{table*}

To study the production of missing $2D$-wave and $1F$-wave states, we first need to calculate the decay width of $\psi(4230) \to X\gamma$ processes. However, for such high-lying charmonium state as $\psi(4230)$, the unquenched effect cannot be ignored. Therefore, we cannot calculate these radiative decay processes simply through the quenched models as introduced in Sec. \ref{Rad}. Instead, in this work, we investigate the production behavior of high-lying charmonium states through the radiative decay processes by hadronic loop mechanism, an approach widely used to study hadronic decays in both the charmonium and bottomonium sectors \cite{Peng:2024xui, Bai:2024lps, Li:2021jjt, Bai:2022cfz, Qian:2023taw, Gao:2024qth}. For the $\eta_c (2D)$, $\chi_{c2} (1F)$, and $\chi_{c3} (1F)$ states, we focus on the specific radiative decay channels of the $\psi (4230)$. The $1^3 F_4$ state is excluded from our study because its high angular momentum strongly suppresses its production from the $e^+e^-$ processes. The decay diagrams allowed for the included processes are presented in Figs. \ref{Figpro}.

As shown in Table \ref{TCP}, the remaining $2D$ and $1F$ states cannot be produced via radiative decays. Alternative production channels, such as the hidden-charm processes $e^+e^- \to \psi \to X\eta$ or $e^+e^- \to \psi \to X\pi\pi$ (where $X$ denotes the missing $2D$ and $1F$ states), could be considered. However, these processes require a center of mass energy exceeding 4.7 GeV, a region where no suitable $\psi$ resonance has been conclusively observed. Due to the current lack of relevant experimental data, we will not perform detailed calculations for these channels. Instead, we will limit our discussion to assessing their feasibility.

\subsection{Hadronic loop mechanism } \label{loop} 

In the framework of the hadronic loop mechanism, the initial state first decays into a $D^{(*)}\bar{D}^{(*)}$ pair, which subsequently transforms into final states through the exchange of a $D^{(*)}$ meson. The decay amplitudes can be expressed as follows

\begin{equation}\label{5.1}
    \mathcal{M}=\int \frac{d^4 q}{(2\pi)^4} \frac{\mathcal{V}_1 \mathcal{V}_2 \mathcal{V}_3}{\mathcal{D}_1\mathcal{D}_2\mathcal{D}_E} \mathcal{F}^2(q^2),
    \end{equation}
where $\mathcal{V}_i (i=1,2,3)$ are interaction vertices, and $\mathcal{D}_i (i=1,2,E)$ denote the corresponding propagators of intermediate charmed mesons. A form factor $\mathcal{F}(q^2)$ is introduced to compensate for the off-shell effect of the exchanged $D^{(*)}$ meson and to depict the structure effect of the interaction vertices. {More importantly, it serves a function similar to Pauli-Villars regularization, preventing divergences in loop integrals. In previous studies \cite{Peng:2024xui, Bai:2024lps, Li:2021jjt, Bai:2022cfz, Qian:2023taw, Gao:2024qth, Chen:2013cpa}, the monopole and dipole form factors were applied in concrete calculations. However, for the processes considered in this work, high orbital states are involved. Therefore, we introduce the dipole form factor to prevent divergences in loop integrals, as a monopole form factor cannot achieve this.} The dipole form factor has the form 

\begin{equation}\label{5.2}
\mathcal{F}(q^2)=\left(\frac{m^2_E-\Lambda^2}{q^2-\Lambda^2}\right)^2,
\end{equation}
where $m_E$ and $q$ represent the mass and four-momentum of the exchanged charmed mesons. The cut-off parameter is defined as $\Lambda = m_E +\alpha \Lambda_{QCD}$, where $\Lambda_{QCD} = 220$ MeV and $\alpha$ is a dimensionless free parameter, typically expected to be of order 1 \cite{Cheng:2004ru}. 

We use the effective Lagrangian approach to describe the diagrams in Fig. \ref{Figpro}. 
The Lagrangians depicting the interaction vertices between charmed mesons and $\psi$ states are provided in \cite{Peng:2024xui}

 \begin{align}\label{5.3}
   \mathcal{L}_{\psi D^{(*)}D^{(*)}}=&i\ g_{\psi DD} \psi^{\mu}(\bar{D} \mathop{\partial_\mu}\limits^{\leftrightarrow} D)  \nonumber\\ & +g_{\psi DD^*}   \varepsilon_{\mu\nu\alpha\lambda}\partial^\mu\psi^{\nu}(D^{*\alpha}\mathop{\partial^\lambda}\limits^{\leftrightarrow}\bar{D}-\bar{D}^{*\lambda} \mathop{\partial^\alpha}\limits^{\leftrightarrow} D ) \nonumber\\ &+i\ g_{\psi D^*D^*}\psi^{\mu}(\bar{D}^{*}_\nu \partial^\nu D^{*}_{\mu} + \partial_\nu \bar{D}^{*}_\mu  D^{*\nu}+ \bar{D}^{*}_\nu \mathop{\partial_\mu}\limits^{\leftrightarrow} D^{*\nu}).
    \end{align}

 The $\mathop{\partial^\mu}\limits^{\leftrightarrow}$ means $\mathop{\partial^\mu}\limits^{\rightarrow}-\mathop{\partial^\mu}\limits^{\leftarrow}$. Then, the effective Lagrangians describing the vertices where $\eta_{c2}(2^1D_2)$ interacts with charmed mesons are \cite{Casalbuoni:1996pg, Li:2021jjt}

\begin{align}\label{5.7}
    \mathcal{L}_{\eta_{c2}(2D)D^{(*)}D^*}=&-2g_{\eta_{c2} DD^*} \psi_2^{\mu\nu} (\bar{D}\mathop{\partial_\mu}\limits^{\leftrightarrow} D^*_\nu+\bar{D}^*_\nu \mathop{\partial_\mu}\limits^{\leftrightarrow} D) \nonumber\\ & +2i\ g_{\eta_{c2} D^*D^*} \partial_\theta \psi_{2\mu\nu} \varepsilon^{\rho\sigma\nu\theta} (\bar{D}^*_\sigma \mathop{\partial^\mu}\limits^{\leftrightarrow} D^{*}_\rho).
 \end{align}

And the interaction Lagrangians describing the $D^{(*)}D^{(*)}\gamma $ vertices are given by \cite{Chen:2013cpa,Duan:2024zuo,Gao:2024qth}

\begin{align}\label{5.8}
\mathcal{L}_{D^{(*)}D^{(*)}\gamma} = & ieA_{\mu}D^-\partial^{\mu}D^+ +\frac{e \  f_{\gamma D D^{*}}}{4} \varepsilon^{\mu\nu\beta\tau} F_{\mu\nu} \nonumber\\
&\times [(\partial_\beta D_{\tau}^{*}-\partial_\tau D_{\beta}^{*}) \bar{D} + (\partial_\tau \bar{D}_{\beta}^{*}-\partial_\beta \bar{D}_{\tau}^{*}) D ] \nonumber\\ &+ ieA^{\mu} ( g^{\tau\beta} D_{\beta}^{*-} \partial_{\tau} D_{\mu}^{*+} - g^{\tau\beta} \partial_{\tau} D_{\mu}^{*-} D_{\beta}^{*+} \nonumber\\ &+ g^{\tau\beta} D_{\beta}^{*-}\mathop{\partial_\mu}\limits^{\leftrightarrow} D_{\tau}^{*+}),
\end{align}
where $e$ represents the unit charge, and the $F^{\mu\nu}$ represents $\partial^\mu A^\nu-\partial^\nu A^\mu$. Here the $D \bar{D}^{*}$ is used as a shorthand that encompasses both $D^+ D^{*-}$ and $D^0 \bar{D}^{*0}$, other charm meson pairs are denoted similarly. We substitute the coupling constant $f_{\gamma DD^*}$ with the $g_{\gamma D^+D^{*-}}$ and $g_{\gamma D^0 \bar{D}^{*0}}$ to represent the different interaction strength between the $\gamma D^+D^{*-}$ and the $\gamma D^0 \bar{D}^{*0}$. The mass of the $\psi(4230)$ lies close to the $D\bar{D}_1(2420)$ threshold. Importantly, a $D\bar{D}_1(2420)$ pair can couple in an $S$-wave to form a state with $J^{PC}=1^{--}$, suggesting a strong coupling between this hadronic channel and the $\psi(4230)$. Recent coupled-channel analyses \cite{Lu:2017yhl, Man:2025zfu} have provided quantitative support for this picture. Consequently, contributions from the $D\bar{D}_1(2420)$ channel should be incorporated in a rigorous treatment of the $\psi(4230) \to \gamma X$ processes. The corresponding interaction Lagrangians are given in Appendix \ref{EL}.

The Lagrangians describing the interactions of other $D$-wave and $F$-wave charmonia with charmed mesons are presented in Appendix \ref{EL}, where the parameters used in our calculations are also listed. For processes involving a photon, gauge invariance requires the amplitudes to satisfy the Ward-Takahashi identity. However, the decay amplitudes listed in Appendix \ref{EL} do not automatically fulfill this condition. To address this issue, we introduce the contact diagrams shown in Fig. 6 \cite{Chen:2013cpa, Cheng:2004ru}. The corresponding interaction Lagrangians for the $\eta_{c2}$ state, describing the $D^* D^{(*)} \gamma$ vertices, can be written as

 \begin{align}\label{5.10}
     \mathcal{L}_{\eta_{c2}(2D) D^* D^{(*)}\gamma}=&-2 i e \ f_{\eta_{c2} DD^*\gamma} A^a \psi_2^{\mu\nu}(\bar{D}\mathop{\partial_\mu}\limits^{\leftrightarrow} \mathop{\partial_a}\limits^{\leftrightarrow} D^*_\nu)  \nonumber\\ & -2i e \ f_{\eta_{c2} DD^*\gamma} A^a \psi_2^{\mu\nu} (\bar{D}^*_\nu \mathop{\partial_\mu}\limits^{\leftrightarrow} \mathop{\partial_a}\limits^{\leftrightarrow} D) \nonumber\\ & -2 e \ g_{\eta_{c2} D^*D^*\gamma}  A^\theta \psi_2^{\mu\nu} \varepsilon_{\rho\sigma\nu\theta} (\bar{D}^{*\sigma} \mathop{\partial_\mu}\limits^{\leftrightarrow} D^{*\rho}).
     \end{align}

Similar to the Eq. (\ref{5.8}), we substitute the coupling constant $f_{\eta_{c2} \gamma DD^*}$ with the $g_{\eta_{c2} \gamma D^+D^{*-}}$ and $g_{\eta_{c2} \gamma D^0 \bar{D}^{*0}}$ to represent the different interaction strength between the $\gamma D^+D^{*-}$ and the $\gamma D^0 \bar{D}^{*0}$.

 For the process $\psi(4230) \to \eta_{c2}(2D) \gamma$, we can write down the concrete amplitudes corresponding to the diagrams in Fig. \ref{Figpro}. For example, with the Feynman rules listed in Appendix \ref{FL}, the amplitude of the first triangle diagram in the left column of Fig. \ref{Figpro} can be obtained as

\begin{align}\label{5.11}
    \mathcal{M}=& -2ie \int \frac{d^4 q}{(2\pi)^4} \ g_{\psi DD^*} g_{\eta_{c2} DD^*} p_1^h g_{\rho\nu} \epsilon^{*\mu\nu}_{\eta_2} \epsilon^{l}_{\psi} \epsilon^{*m}_{\gamma}  \nonumber\\ & \times  \varepsilon_{hl\lambda j}  (q2_\mu-q_{\mu})(q_m+q_{1m}) (q_{1}^j-q_{2}^j) \nonumber\\ & \times (\frac{q_2^\lambda q_{2\rho}}{m^2_{q_2}}-g^{\lambda}_\rho) \frac{1}{q_1^2-m^2_D} \frac{1}{q_2^2-m^2_{D^*}} \frac{1}{q^2-m^2_D}.
    \end{align}

For the contact amplitudes, we replace the vertices described in Eqs. (\ref{5.7})-(\ref{5.8}) with the interaction vertices in Eq. (\ref{5.10}), then we can obtain the contact amplitudes similarly. All the Lagrangians listed in Appendix \ref{EL} can be easily translated into amplitudes in the same way. The total amplitude for $\psi(4230) \to \eta_{c2} (2^1D_2) \gamma$ can then be expressed as

\begin{align}\label{5.13}
\mathcal{M}_{\text{Tot}}=\sum_{\text{all tri}}\mathcal{M}_{\text{tri}} + \sum_{\text{all con}} \mathcal{M}_{\text{con}}.
\end{align}

The $\mathcal{M}_{\text{tri}}$ represent amplitudes of triangle diagrams in Fig. \ref{Figpro}, and the $\mathcal{M}_{\text{con}}$ represent amplitudes of contact diagrams in Fig. \ref{Figpro}. For the radiative decay processes, the diagrams always include an external photon line, therefore the total amplitudes can be written as $\mathcal{M}^\mu_{\text{Tot}} \epsilon^*_{\gamma\mu}$. To satisfy the Ward-Takahashi identity, the total amplitude must satisfy the condition that $\mathcal{M}^\mu_{\text{Tot}} p_{2\mu} = 0$ when the photon polarization vector $\epsilon^*_{\gamma\mu}$ is replaced by the photon four-momentum $p_{2\mu}$. 

We will take the $\eta_c (2D)$ state as an example. If we replace the $\epsilon^*_{\gamma\mu}$ with $p_{2\mu}$, then we obtain 

\begin{align}\label{5.15}
    \sum\mathcal{M}_{con} \propto & \  (p_{2}^\mu \varepsilon^{vabc}p_{2a} \epsilon^*_{\gamma b} \epsilon_{\psi c} \epsilon^*_{\eta_c \mu\nu} + p_{2}^\mu \varepsilon^{vabc}p_{3a} \epsilon^*_{\gamma b} \epsilon_{\psi c} \epsilon^*_{\eta_c \mu\nu}).
\end{align}
More details can be found in the Appendix \ref{GI}. Having obtained the total amplitudes, we proceed to calculate the production widths of the $D$-wave and $F$-wave states using the following formula
\begin{align}\label{5.16}
    \Gamma= \frac{1}{2J+1} \frac{|\vec{p}|}{8\pi m_{\psi}^2} |\mathcal{M}^{\rm Total}|^2,
\end{align}
where $\vec{p}$ represents the center-of-mass momentum of the final state, and the factor of 1/$(2J+1)$ accounts for the averaging over the polarization of the initial state. The partial width of the production processes of other final states can be calculated in a similar manner.

\subsection{ Numerical results of production processes } \label{Result}

In this section, we will exhibit the numerical results of chosen radiative decay processes and predict the possibility of the observation of these states in radiative decay processes. To calculate these diagrams listed in Fig. \ref{Figpro}, we need to determine the coupling constants in the amplitudes. The specific coupling constants are listed in Appendix \ref{EL}.

With the coupling constants determined, we can proceed to evaluate the decay widths and branching ratios. However, our model still contains a free parameter, $\alpha$, which enters through the cutoff $\Lambda$ in the dipole form factor. Since $\Lambda$ should be comparable to the physical mass of the exchanged meson, $\alpha$ is typically expected to be of order 1. Additionally, the dipole form factor behaves like a propagator of mass $\Lambda$, introducing an extra branch cut in the loop integral when two intermediate particles can go on-shell simultaneously—a condition that occurs if $\Lambda + m_{D^{(*)}} = m_{\eta_{c2}(2D)}$. In this work, we take $\alpha$ in the range $1.5$–$2$ for the $\psi(4230) \to X\gamma$ transitions when presenting numerical results.

{Indeed, our results are sensitive to the choice of the $\alpha$ value. In practice, experimental data are needed to constrain its range. In this work, based on previous studies, we expect $\alpha$ to be of order $\mathcal{O}(1)$. We therefore present our results for $\alpha$ values ranging from 1.5 to 2, thereby allowing us to identify which processes are more likely to be accessible in future experiments. In particular, we hope to encourage experimental interest in exploring such processes.}


The widths and branching ratios of chosen processes are presented in Fig. \ref{Result}. We notice that the branching ratios of $\psi(4230) \to \chi_{c2}(1F) \gamma$ and $\psi(4230) \to \chi_{c3}(1F) \gamma$ processes are relatively much larger than the branching ratio of $\psi(4230) \to \eta_{c2}(2D) \gamma$. 

The partial width for $\chi_{c3}(1F)$ production varies from approximately 5 keV to 50 keV as $\alpha$ increases from 1.5 to 2, corresponding to a branching ratio on the order of $10^{-5}$. In contrast, the $\chi_{c2}(1F)$ state exhibits a larger width, reaching about 180 keV for $\alpha=2$, with its branching ratio around $10^{-4}$. This marked difference arises from their distinct production mechanisms: $\chi_{c2}(1F)$ is produced via an $S$-wave, while $\chi_{c3}(1F)$ proceeds through a $D$-wave, which is typically more suppressed.  Furthermore, the ratio $\Gamma_{\gamma\chi_{c2}} / \Gamma_{\gamma\chi_{c3}}$ remains stable at approximately 5 across the considered $\alpha$ range. This indicates that the $\chi_{c2}(1F)$ state should be considerably easier to observe in radiative decay channels. 

However, for the $\eta_{c2}(2D)$ state, we find its partial width is exceptionally small, on the order of $10^{-7}$ MeV (i.e., $10^{-4}$ keV) for $\alpha$ in the range 1.5–2, corresponding to a branching ratio of $\mathcal{O}(10^{-9})$. Furthermore, this width exhibits only a mild dependence on $\alpha$, in sharp contrast to the strong sensitivity observed for the $\chi_{c2}(1F)$ and $\chi_{c3}(1F)$ states. The extreme smallness of this width renders the $\psi(4230) \to \gamma \eta_{c2}(2D)$ channel negligible among all decay modes of the $\psi(4230)$. This suppression can be attributed to two main factors. First, the relatively high mass of the $\eta_{c2}(2D)$ final state leads to a reduced phase space. However, phase-space considerations alone cannot account for the observed orders-of-magnitude suppression. A more fundamental reason lies in the heavy-quark spin symmetry (HQSS): transitions between a vector (spin-triplet) charmonium like the $\psi(4230)$ and a tensor spin-singlet state like the $\eta_{c2}(2D)$ are strongly suppressed \cite{BESIII:2025bce}. The HQSS-breaking interactions that enable such spin-flip transitions are inherently small, leading to the tiny width observed.

\begin{figure*}[htbp]
	\centering
	\begin{tabular}{c}
		\includegraphics[width=1\textwidth]{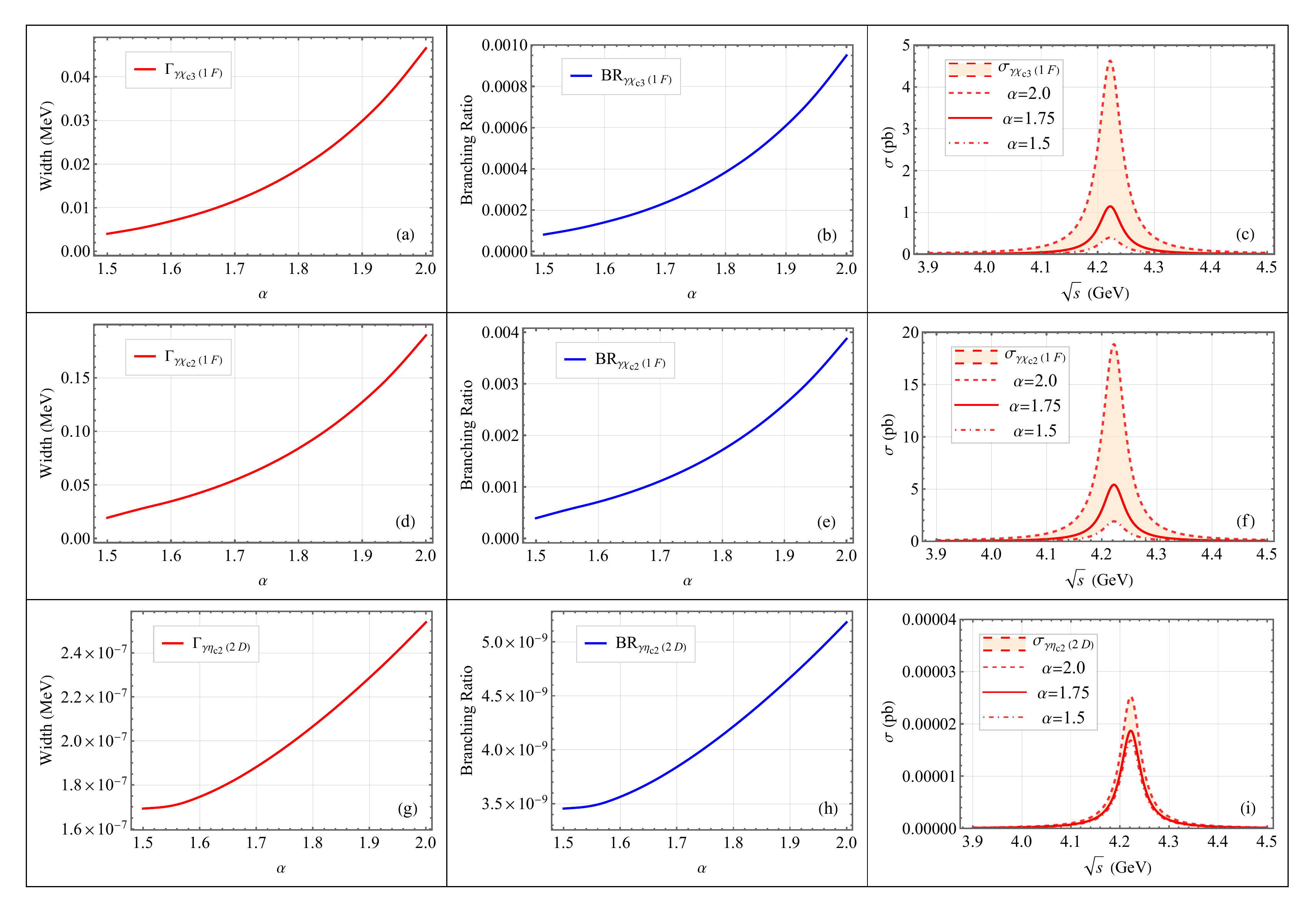}
	\end{tabular}
    \captionsetup{justification=raggedright, singlelinecheck=false}
	\caption{The numerical results for the process $e^+e^- \to \psi(4230) \to \gamma X$. The left, middle, and right columns depict the partial widths, branching ratios (among all $\psi(4230)$ decay modes), and production cross sections, respectively, for each final state $X$. In the right column, the cross sections are plotted for three typical values of the parameter $\alpha$: 1.5, 1.75, and 2, distinguished by different line styles.}
	\label{Result}
\end{figure*}

After we obtain the partial width of these production processes, we are able to calculate their scattering cross section $\sigma(X\gamma)$. To estimate the $\sigma(X\gamma)$, the cross sections for $e^+e^- \to \gamma X$  can be expressed as

\begin{align}\label{5.17}
    \sigma(e^+ e^- \to \psi(4230) \to \gamma X)= \frac{12\pi\Gamma^{e^+e^-}_{\psi}\Gamma_\psi^{\gamma X}}{|s-m_\psi^2+im_\psi\Gamma_\psi|^2},
\end{align}
where $s$ is the square of the center-of-mass energy, $\Gamma^{e^+e^-}_{\psi}$ denotes the dilepton decay width of $\psi(4230)$ and $\Gamma_\psi^{\gamma X}$ represents the partial width for the $\psi(4230) \to \gamma X$ processes, while the $X$ denotes the $D$-wave or $F$-wave charmonium states under study. Here, $m_\psi$ and $\Gamma_\psi$ refer to the mass and total width of the charmonium state $\psi(4230)$, respectively. The dilepton width $\Gamma^{e^+e^-}_{\psi}$ is taken from the theoretical calculation in Ref. \cite{Wang:2019mhs}, which gives a value of 0.290 keV. Using the partial widths $\Gamma_\psi^{\gamma X}$ calculated in this work, we can estimate the production cross sections $\sigma(X\gamma)$. The predicted cross sections as a function of center-of-mass energy for $\sqrt{s}$ ranging from 3.9 GeV to 4.5 GeV are illustrated in Fig. \ref{Result}.

The calculated cross section for producing $\chi_{c2}(1F)$ is the largest in our study, reaching $\sim 20\text{ pb}$ at $\alpha = 2$. This prominence is consistent with its dominant $S$-wave production mechanism, as discussed earlier. We find this cross section is highly sensitive to the parameter $\alpha$, with a marked increase for $\alpha > 1.7$. To quantify this model dependence, Fig. \ref{Result} displays the results for three benchmark values: $\alpha = 1.5$, $1.75$, and $2$, corresponding to a variation in $\sigma(e^+e^- \to \gamma\chi_{c2}(1F))$ from about $2\text{ pb}$ to $20\text{ pb}$. This range provides an estimate of the theoretical uncertainty associated with the form-factor cutoff. Our predicted cross sections suggest that the process $e^+e^- \to \gamma \chi_{c2}(1F)$ could yield a detectable number of events in existing or near-future datasets. Therefore, searching for the $\chi_{c2}(1F)$ state via $e^+ e^- \to \psi(4230) \to \gamma \chi_{c2}(1F)$ appears to be a promising avenue.

For the $\chi_{c3}(1F)$ state, the production cross section $\sigma(\gamma\chi_{c3}(1F))$ is smaller than that of $\chi_{c2}(1F)$, consistent with its suppressed $D$-wave production. Quantitatively, it varies from approximately $0.5$ pb to $5$ pb as $\alpha$ increases from $1.5$ to $2$, with a value of about $1$ pb at $\alpha = 1.75$. This order-of-magnitude suppression relative to $\chi_{c2}(1F)$ implies that observing $\chi_{c3}(1F)$ in $e^+e^- \to \psi(4230) \to \gamma \chi_{c3}(1F)$ with current integrated luminosity is challenging. However, the planned upgrades of BESIII and the future STCF, which promise a substantial increase in luminosity, will provide an ideal opportunity to search for such suppressed channels. Our calculation indicates that $\chi_{c3}(1F)$ is well within the reach of these next-generation experiments.

The predicted production cross section for the $\eta_{c2}(2D)$ state via $\psi(4230)$ radiative decay is exceedingly small, $\sigma(\gamma\eta_{c2}(2D)) \sim 2 \times 10^{-5}\ \text{pb}$, and shows negligible variation over the chosen range of $\alpha$. A cross section of this magnitude renders the channel $e^+e^- \to \psi(4230) \to \gamma \eta_{c2}(2D)$ effectively unobservable with foreseeable integrated luminosities of BEPC as listed above. Consequently, the discovery of the $\eta_{c2}(2D)$ likely relies on alternative production mechanisms. A prime example is its candidate state, the $X(4160)$, which was first observed not via radiative transitions but in the double charmonium production
process $e^+e^- \to J/\psi D^{(*)} \bar{D}^{(*)}$ at $\sqrt{s} \sim 10.6$ GeV by Belle \cite{Belle:2007woe}. While challenging, this highlights that high-energy $e^+e^-$ scans or dedicated $B$-meson decay analyses (as with the later LHCb observation \cite{LHCb:2021uow}) are more viable paths. Future experiments like the Belle II and LHCb, designed to operate at higher center-of-mass energies with high luminosity, are well-suited to systematically search for such states in both exclusive and inclusive channels.

\subsection{ The production of other missing charmonia } \label{Other} 

The production of the $\chi_{c4}(1F)$ state faces significant suppression from two factors: it is also a $D$-wave process, and it carries a high total angular momentum of $J=4$. This combination makes its expected production cross section exceedingly small. Consequently, we do not calculate its production channels in this work. It is worth noting, however, that the next generation of high-intensity colliders, such as the STCF, are designed to explore precisely such rare and highly suppressed processes, potentially opening a window to these extreme angular momentum states.

The production processes for the remaining missing states are not explicitly calculated in this work. Instead, we assess their potential observability. These states cannot be produced via radiative transitions of the $\psi(4230)$ ($e^+e^- \to \gamma X$) due to quantum number constraints and kinematic thresholds. Their primary production mechanism at $e^+e^-$ colliders would be through hidden-charm decays of higher-mass charmonia, such as $e^+e^- \to \psi \to \eta X$ or $e^+e^- \to \psi \to \omega X$. However, these processes require the initial charmonium mass to exceed significant thresholds, which should be at least approximately 4.6 GeV for $\eta X$ and 4.8 GeV for $\omega X$ final states. As we discussed, the experimental data in this high-mass region ($\gtrsim$ 4.6 GeV) are sparse, and the charmonium spectrum itself is poorly established here. 

Alternatively, these high-lying states could be sought in $B$-meson decay channels, as studied by the LHCb experiment. However, making precise theoretical predictions for such processes is challenging due to their non-perturbative nature and the current lack of experimental data to tightly constrain the models. A pertinent example is the $X(4160)$, which is discovered in $B^+ \to J/\psi \phi K^+$ process by LHCb \cite{LHCb:2021uow}. The latter observation highlights the potential of $B$-decay channels to access charmonium-like states that may be suppressed or inaccessible in direct $e^+e^-$ annihilation. The different production mechanism and kinematic environment in $B$ decays offer a complementary probe, making them a promising avenue for discovering some of the missing high-lying charmonia.

Given these theoretical and experimental uncertainties, a targeted search remains difficult at present. Looking ahead, the next generation of high-luminosity facilities offers a more promising path. The STCF will provide a pristine $e^+e^-$ environment ideally suited for systematic scans of the charmonium spectrum above 4.6 GeV. Concurrently, upgrades to the LHCb experiment will deliver vastly increased statistics in $B$-decay channels. Together, these advances are designed to explore this uncharted energy region, potentially enabling the discovery of these high-spin, high-mass charmonia in the coming years.

\section{DISCUSSIONs AND conclusions}

Our understanding of high-lying states in the charmonium family remains limited. With the recent observation of charmonium states above 4 GeV, we take the opportunity to systematically investigate these high-lying states. This study includes a mass spectrum analysis using the MGI model, calculations of two-body strong decays within the QPC model, and an analysis of radiative decays. This theoretical framework enables us to extract resonance parameters for the discussed charmonium states, offering valuable guidance for future research in heavy hadron spectroscopy.

In this work, we compute the mass spectrum of $2D$ and $1F$ charmonium states by incorporating unquenched effects via a screening potential. We also provide a detailed analysis of their strong decay properties, including key decay channels and partial widths that are essential for identifying possible charmonium candidates in forthcoming experiments. Additionally, we examine radiative decay behaviors and highlight the dominant radiative channels for the states under consideration.

Among $D$-wave states, the $1D$ triplet and the $2^{3}D_{1}$ state have already been observed. It is therefore natural to explore possible experimental avenues for detecting the $2^{3}D_{2}$. Given that these states are relatively narrow and decay predominantly into $D\bar{D}^{*}$ and $D^{*}\bar{D}^{*}$, we suggest experimental searches in the $D\bar{D}^{*}$ and $D^{*}\bar{D}^{*}$ channels. As a reference, the $1^{3}D_{2}$ state $\psi_{2}(3823)$ was previously discovered in $B \to \chi_{c1}\gamma K$ and $B^{+} \to J/\psi\pi^{+}\pi^{-} K^{+}$ processes \cite{Belle:2013ewt,LHCb:2020fvo}. Similarly, the $2^{3}D_{2}$ and $2^{3}D_{3}$ states could be sought in $B \to D^{*}DK$ and $B \to D^{*}D^{*}K$ processes in future experiments by LHCb and Belle II. However, a recent LHCb analysis of $B \to D^{*}DK$~\cite{LHCb:2024vfz} found no evidence for these states. We hope that more precise future data will help uncover them.

Besides $B$-meson decays, strong decays of higher excited $1^{--}$ charmonia produced directly in $e^{+}e^{-}$ annihilation also offer promising pathways to search for the $2^{3}D_{2}$ and $2^{3}D_{3}$ states. For example, processes such as $e^{+}e^{-} \to \psi \to \eta \psi_{2/3} \to \eta D^{*}\bar{D}^{(*)}$ could provide valuable detection opportunities. A rough estimate suggests that the mass of the intermediate $\psi$ state would likely exceed 4.7 GeV. Moreover, our results show a mass splitting between the $2^{3}D_{2}$ and $2^{3}D_{3}$ states of about 7 MeV. Distinguishing between them will require high statistics to determine their spins precisely, posing significant challenges for experiments at the upgraded BESIII and Belle II.

The successful identification of the $1D$ triplet encourages us to examine production mechanisms for states with higher orbital angular momentum, such as the $1F$ triplet. Although the masses of the $1F$ triplet are close, their dominant decay channels differ. Based on the open-charm branching ratios shown in Table~\ref{Ftable}, we propose that the $1^{3}F_{2}$ state could be probed in the $D\bar{D}$ channel, whereas the $1^{3}F_{4}$ state could be studied in the $D^{*}\bar{D}^{*}$ channel. Since both the $1^{3}F_{2}$ and $1^{3}F_{3}$ states have sizable branching ratios into $D\bar{D}^{*}$, distinguishing them in the $D\bar{D}^{*}$ channel remains challenging.

Analogous to the $2^{3}D_{2}$ and $2^{3}D_{3}$ cases, the $1F$ charmonium triplet can be investigated through two main approaches: (1) $B$-meson decays and (2) hidden-charm decays of higher excited $1^{--}$ charmonia, both requiring high energy. Furthermore, the $C$-even nature of the $1F$ triplet allows production via radiative transitions from excited vector charmonia with masses above 4.2 GeV. In particular, the $1^{3}F_{2}$ state with $J^{PC}=2^{++}$ is a prime candidate for $S$-wave-dominated radiative transitions from excited $1^{--}$ charmonia. We therefore encourage experimental collaborations such as BESIII and Belle II to consider the process $e^{+}e^{-} \to \gamma D\bar{D}$ in future searches for the $1^{3}F_{2}$ state.

In summary, although observing high-lying charmonium states is challenging, several promising avenues exist for detecting these elusive states in the future, including $B \to D^{*}D^{*}K$, $e^{+}e^{-} \to \eta D^{*}\bar{D}^{(*)}$, and $e^{+}e^{-} \to \gamma D\bar{D}^{(*)}$.

We are entering a new era of high-precision hadron spectroscopy, driven by upgrades to the Large Hadron Collider (LHC), the operation of SuperKEKB, improvements at the Beijing Electron–Positron Collider (BEPCII‑U), and upcoming facilities such as the STCF. The theoretical insights presented in this work on high-lying charmonium mesonic states may provide important guidance for experimental studies at BESIII, Belle II, and LHCb.

\begin{acknowledgments}
This work is supported by the National Natural Science Foundation of China under Grant Nos. 12335001 and 12247101, the `111 Center' under Grant No. B20063, the Natural Science Foundation of Gansu Province (No. 22JR5RA389, No. 25JRRA799), the Talent Scientific Fund of Lanzhou University, the fundamental Research Funds for the Central Universities (No. lzujbky-2023-stlt01), the project for top-notch innovative talents of Gansu province, and Lanzhou City High-Level Talent Funding.
\end{acknowledgments}
 
\begin{appendix}

\begin{widetext}

\section{Effective Lagrangians} \label{EL}

The interactions of $S$-wave or $D$-wave charmonia and charmed mesons are listed in Ref. \cite{Li:2021jjt}, whose Lagrangians can be written as
 \begin{align}\label{A1}
\mathcal{L}_{S}=i\ g_S \text{Tr}[S_{Q\bar{Q}} \bar{H}^{\bar{Q}q} \gamma^\mu\mathop{\partial_\mu}\limits^{\leftrightarrow}  \bar{H}^{Q\bar{q}}],
\\  \mathcal{L}_{D}=i\ g_D \text{Tr}[D^{\mu\nu}_{Q\bar{Q}} \bar{H}^{\bar{Q}q} \mathop{\partial_\mu}\limits^{\leftrightarrow} \gamma_\nu \bar{H}^{Q\bar{q}}].
    \end{align}
    
Besides, Lagrangians of interactions between charmed mesons and $F$-wave charmonia are given by

\begin{align}\label{A3}
    \mathcal{L}_{F}=i\ g_F \text{Tr}[F^{\mu\nu\theta}_{Q\bar{Q}} \bar{H}^{\bar{Q}q} \mathop{\partial_\mu}\limits^{\leftrightarrow} \mathop{\partial_\theta}\limits^{\leftrightarrow} \gamma_\nu \bar{H}^{Q\bar{q}}]. 
    \end{align}

The $\mathop{\partial^\mu}\limits^{\leftrightarrow}$ means $\mathop{\partial^\mu}\limits^{\rightarrow}-\mathop{\partial^\mu}\limits^{\leftarrow}$. The $\bar{H}$ is related to the doublet field of charmed meson field $H=(\frac{1+\slashed{v}}{2})(D^{*\mu}\gamma_{\mu}+iD\gamma_5)$ by $\bar{H}=\gamma_0 H^\dagger \gamma_0$. The $S^{Q\bar{Q}}$, $D_{\mu\nu}^{Q\bar{Q}}$, and $F_{\mu\nu\theta}^{Q\bar{Q}}$ represent the $S$-wave, $D$-wave and $F$-wave multiplets, which can be written as \cite{Casalbuoni:1996pg}

\begin{align}\label{A4}
    S_{Q\bar{Q}}=&\frac{1+\slashed{v}}{2} (\psi^\mu\gamma_\mu-\eta_c\gamma_5)\frac{1-\slashed{v}}{2},
\\
 \nonumber\\
    D_{Q\bar{Q}}^{\mu\nu}=&\frac{1+\slashed{v}}{2} (\psi_3^{\mu\nu\alpha}\gamma_\alpha+\frac{1}{\sqrt{6}}(\varepsilon^{\mu\alpha\beta\gamma}v_\alpha\gamma_\beta\psi_{2\gamma}^\nu+\varepsilon^{\nu\alpha\beta\gamma}v_\alpha\gamma_\beta\psi_{2\gamma}^\mu) \nonumber\\& +\frac{\sqrt{15}}{10}((\gamma^\mu-v^\mu)\psi_1^\nu+(\gamma^\nu-v^\nu)\psi_1^\mu) -\frac{1}{\sqrt{15}}(g^{\mu\nu}-v^\mu v^\nu)\gamma_\alpha\psi_1^\alpha+\eta_{c2}^{\mu\nu}\gamma_5)\frac{1-\slashed{v}}{2},
\\
 \nonumber\\
    F_{Q\bar{Q}}^{\mu\nu\theta}=&\frac{1+\slashed{v}}{2} (\chi_{c4}^{\mu\nu\theta\alpha}\gamma_\alpha +\frac{1}{\sqrt{12}}(\varepsilon^{\mu\alpha\beta\gamma}v_\alpha\gamma_\beta\chi_{3\gamma}^{\nu\theta} +\varepsilon^{\nu\alpha\beta\gamma}v_\alpha\gamma_\beta\chi_{c3\gamma}^{\mu\theta}+\varepsilon^{\theta\alpha\beta\gamma}v_\alpha\gamma_\beta\chi_{c3\gamma}^{\mu\nu}) \nonumber\\& +\frac{\sqrt{35}}{21}((\gamma^\mu-v^\mu)\chi_{c2}^{\nu\theta}+(\gamma^\nu-v^\nu)\chi_{c2}^{\mu\theta}+(\gamma^\theta-v^\theta)\chi_{c2}^{\mu\nu}) \nonumber\\&-\frac{2}{3\sqrt{35}}((g^{\mu\nu}-v^\mu v^\nu)\gamma_\alpha\chi_{c2}^{\alpha\theta}+(g^{\nu\theta}-v^\nu v^\theta)\gamma_\alpha\chi_{c2}^{\alpha\mu} +(g^{\mu\theta}-v^\mu v^\theta)\gamma_\alpha\chi_{c2}^{\alpha\nu}) +h_{c3}^{\mu\nu\theta}\gamma_5)\frac{1-\slashed{v}}{2}.
\end{align}

The $\chi_{c4}$ and $\psi_1$ terms are not involved in our work. The Lagrangians describing the interaction of charmed mesons with a photon can be written as

\begin{align}\label{A11}
\mathcal{L}_{D^{(*)}D^{(*)}\gamma} = & ieA_{\mu}D^-\partial^{\mu}D^+  +\frac{e \, f_{D \bar{D}^{*}\gamma}}{4} \varepsilon^{\mu\nu\beta\tau} F_{\mu\nu}[(\partial_\beta D_{\tau}^{*}-\partial_\tau D_{\beta}^{*}) \bar{D} + (\partial_\tau \bar{D}_{\beta}^{*}-\partial_\beta \bar{D}_{\tau}^{*}) D ]  \nonumber\\
&+ ieA^{\mu} ( g^{\tau\beta} D_{\beta}^{*-} \partial_{\tau} D_{\mu}^{*+} - g^{\tau\beta} \partial_{\tau} D_{\mu}^{*-} D_{\beta}^{*+} + g^{\tau\beta} D_{\beta}^{*-}\mathop{\partial_\mu}\limits^{\leftrightarrow} D_{\tau}^{*+}).
\end{align}

The effective Lagrangians depicting the interaction vertices between charmonium and a pair of S-wave charmed mesons in our work are listed as follows

 \begin{align}\label{A6}
     \mathcal{L}_{\eta_{c2}(2D)}=&-2g_{\eta_{c2} DD^*} \psi_2^{\mu\nu} (\bar{D}\mathop{\partial_\mu}\limits^{\leftrightarrow} D^*_\nu+\bar{D}^*_\nu \mathop{\partial_\mu}\limits^{\leftrightarrow} D) +2i\ g_{\eta_{c2} D^*D^*} \partial_\theta \psi_{2\mu\nu} \varepsilon^{\rho\sigma\nu\theta} (\bar{D}^*_\sigma \mathop{\partial^\mu}\limits^{\leftrightarrow} D^{*}_\rho),
\end{align}

 \begin{align}\label{A8}
     \mathcal{L}_{\chi_{c2}(1F)}=& 2i\sqrt{\frac{7}{5}} g_{\chi_{c2}DD}\psi_2^{\mu\nu}(\bar{D} \mathop{\partial_\mu}\limits^{\leftrightarrow} \mathop{\partial_\nu}\limits^{\leftrightarrow}D)
     \nonumber\\&+\frac{2}{3\sqrt{35}}g_{\chi_{c2}DD^*} \partial^\chi \psi_2^{\mu\nu}(9D^{*\rho} \mathop{\partial_\mu}\limits^{\leftrightarrow} \mathop{\partial^\gamma}\limits^{\leftrightarrow} \bar{D}\varepsilon_{\rho\gamma\chi\nu}+5D^{*\rho} \mathop{\partial_\nu}\limits^{\leftrightarrow}\mathop{\partial^\gamma}\limits^{\leftrightarrow}\bar{D}\varepsilon_{\rho\gamma\chi\mu}+2\varepsilon_{\rho\chi\mu\nu}D^{*\rho} \mathop{\partial_\gamma}\limits^{\leftrightarrow}\mathop{\partial^\gamma}\limits^{\leftrightarrow}\bar{D})\nonumber\\&-\frac{2}{3\sqrt{35}}g_{\chi_{c2}DD^*} \partial^\chi \psi_2^{\mu\nu}(9\bar{D}^{*\sigma} \mathop{\partial_\mu}\limits^{\leftrightarrow} \mathop{\partial^\gamma}\limits^{\leftrightarrow} D\varepsilon_{\sigma\gamma\chi\nu}+5\bar{D}^{*\sigma} \mathop{\partial_\nu}\limits^{\leftrightarrow}\mathop{\partial^\gamma}\limits^{\leftrightarrow}D\varepsilon_{\sigma\gamma\chi\mu}+2\varepsilon_{\sigma\chi\mu\nu}\bar{D}^{*\sigma}\mathop{\partial_\gamma}\limits^{\leftrightarrow}\mathop{\partial^\gamma}\limits^{\leftrightarrow}D) \nonumber\\ &-\frac{2i}{3\sqrt{35}} g_{\chi_{c2}D^*D^*} \psi_2^{\mu\nu}(11\bar{D}^*_\sigma \mathop{\partial_\mu}\limits^{\leftrightarrow} \mathop{\partial_\nu}\limits^{\leftrightarrow}D^{*}_\rho g^{\rho\sigma} +5\bar{D}^*_\sigma \mathop{\partial^\rho}\limits^{\leftrightarrow} \mathop{\partial_\nu}\limits^{\leftrightarrow} D^{*}_\rho g^{\sigma}_{\mu} +5\bar{D}^*_\sigma \mathop{\partial^\sigma}\limits^{\leftrightarrow} \mathop{\partial_\nu}\limits^{\leftrightarrow} D^{*}_\rho g^{\rho}_{\mu} \nonumber\\& +\bar{D}^*_\sigma \mathop{\partial^\rho}\limits^{\leftrightarrow} \mathop{\partial_\mu}\limits^{\leftrightarrow} D^{*}_\rho g^{\sigma}_{\nu}+\bar{D}^*_\sigma \mathop{\partial^\sigma}\limits^{\leftrightarrow} \mathop{\partial_\mu}\limits^{\leftrightarrow} D^{*}_\rho g^{\rho}_\nu +2\bar{D}^*_\sigma \mathop{\partial^\gamma}\limits^{\leftrightarrow} \mathop{\partial_\gamma}\limits^{\leftrightarrow} D^{*}_\rho g^{\sigma}_\mu g^{\rho}_\nu+2\bar{D}^*_\sigma \mathop{\partial^\gamma}\limits^{\leftrightarrow} \mathop{\partial_\gamma}\limits^{\leftrightarrow} D^{*}_\rho g^{\rho}_\mu g^{\sigma}_\nu),
\end{align}

 \begin{align}\label{A9}
    \mathcal{L}_{\chi_{c3}(1F)}= &
     i\sqrt{3}g_{\chi_{c3}DD^*} \psi_3^{\mu\nu\theta}(\bar{D} \mathop{\partial_\mu}\limits^{\leftrightarrow} \mathop{\partial_\nu}\limits^{\leftrightarrow} D^*_\theta) +i\sqrt{3}g_{\chi_{c3}DD^*} \psi_3^{\mu\nu\theta} (D \mathop{\partial_\mu}\limits^{\leftrightarrow} \mathop{\partial_\nu}\limits^{\leftrightarrow} \bar{D}^*_\theta)
    \nonumber\\&-\sqrt{3}g_{\chi_{c3}D^*D^*} \partial^\chi \psi_3^{\mu\nu\theta}(\bar{D}^{*\sigma} \mathop{\partial_\mu}\limits^{\leftrightarrow} \mathop{\partial^\xi}\limits^{\leftrightarrow}D^{*\rho})  ( g_{\rho\sigma} \varepsilon_{\xi\chi\nu\theta}-g_{\sigma\nu}\varepsilon_{\rho\xi\chi\theta}-g_{\rho\nu} \varepsilon_{\sigma\xi\chi\theta}),
\end{align}

 \begin{align}\label{A10}
   \mathcal{L}_{\psi}=&i\ g_{\psi DD} \psi^{\mu}(\bar{D} \mathop{\partial_\mu}\limits^{\leftrightarrow} D) +g_{\psi DD^*}   \varepsilon_{\mu\nu\alpha\lambda}\partial^\mu\psi^{\nu}(D^{*\alpha}\mathop{\partial^\lambda}\limits^{\leftrightarrow}\bar{D}-\bar{D}^{*\lambda} \mathop{\partial^\alpha}\limits^{\leftrightarrow} D ) \nonumber\\ &+i\ g_{\psi D^*D^*}\psi^{\mu}(\bar{D}^{*}_\nu \partial^\nu D^{*}_{\mu} + \partial_\nu \bar{D}^{*}_\mu  D^{*\nu}+ \bar{D}^{*}_\nu \mathop{\partial_\mu}\limits^{\leftrightarrow} D^{*\nu}).
    \end{align}

Besides, the effective Lagrangians including the interactions of $D_1 (2420)$ and other mesons or photon are listed as follows

\begin{align}\label{A14}
    \mathcal{L}_{\psi DD_1}=&i\ g_{\psi DD1} \psi_{\mu}(\bar{D}  D_1^\mu-D\bar{D}_1^\mu), \nonumber\\
     \mathcal{L}_{D_1D^{(*)} \gamma}=&i\ e f_{DD1 \gamma} A_\mu (D\bar{D}_1^\mu-\bar{D}  D_1^\mu) + i\ e f_{D^*D1 \gamma} \varepsilon_{\mu\nu\alpha\beta}  F^{\mu\nu} (D^{*\alpha}\bar{D}_1^{\beta}-\bar{D}^{*\alpha}  D_1^\beta).
 \end{align}

where the generic coupling constants $f_{DD_1\gamma}$ and $f_{D^*D_1\gamma}$ should be replaced by the specific ones $g_{D^{0}_1D^0 \gamma}$ and $g_{D_1^{+} D^-\gamma}$ (or $g_{D^{0}_1 \bar{D}^{*0} \gamma}$ and $g_{D_1^{+} D^{*-}\gamma}$) to distinguish the interaction strengths between the neutral ($\gamma D^0 \bar{D}^{0}$ or $\gamma D^0 \bar{D}^{*0}$) and charged ($\gamma D^+ D^{-}$ or $\gamma D^+ D^{*-}$) final-state channels, respectively.

In gauge theory, invariance under a local $U(1)$ transformation requires the replacement of the ordinary derivative $\partial_\mu$ with the gauge-covariant derivative $D_\mu = \partial_\mu + ieA_\mu$. The contact terms are necessary to restore gauge invariance in the amplitude and can be systematically obtained by substituting one factor of $\partial_\mu$ with $ieA_\mu$ in the relevant interaction Lagrangian. This replacement ensures that the full amplitude transforms covariantly. The resulting effective Lagrangians for the contact terms are given as follows:

 \begin{align}\label{A11}
     \mathcal{L}_{D D^{(*)} \eta_{c2}(2D)\gamma}=& -2ieg_{\eta_{c2} DD^* \gamma} \psi_2^{\mu\nu}  A^a (\bar{D}\mathop{\partial_\mu}\limits^{\leftrightarrow} \mathop{\partial_a}\limits^{\leftrightarrow} D^*_\nu) -2ieg_{\eta_{c2} DD^* \gamma} \psi_2^{\mu\nu} A^a (\bar{D}^*_\nu \mathop{\partial_\mu}\limits^{\leftrightarrow} \mathop{\partial_a}\limits^{\leftrightarrow} D) \nonumber\\ & -2e\ g_{\eta_{c2} D^*D^* \gamma} A_\theta \psi_{2\mu\nu} \varepsilon^{\rho\sigma\nu\theta} (\bar{D}^*_\sigma \mathop{\partial^\mu}\limits^{\leftrightarrow} D^{*}_\rho),
     \end{align}

 \begin{align}\label{A12}
     \mathcal{L}_{D^{(*)}D_{(1)}^{(*)}\chi_{c2}(1F) \gamma}=& -2e\sqrt{\frac{7}{5}} g_{\chi_{c2}DD \gamma } A_a \partial^a \psi_2^{\mu\nu}(\bar{D} \mathop{\partial_\mu}\limits^{\leftrightarrow} \mathop{\partial_\nu}\limits^{\leftrightarrow}D)
     \nonumber\\&+ie\frac{2}{3\sqrt{35}}g_{\chi_{c2}DD^* \gamma}  A_b \partial^b  \partial^\chi \psi_2^{\mu\nu}(9D^{*\rho} \mathop{\partial_\mu}\limits^{\leftrightarrow} \mathop{\partial^\gamma}\limits^{\leftrightarrow} \bar{D}\varepsilon_{\rho\gamma\chi\nu} +5D^{*\rho} \mathop{\partial_\nu}\limits^{\leftrightarrow}\mathop{\partial^\gamma}\limits^{\leftrightarrow}\bar{D}\varepsilon_{\rho\gamma\chi\mu}+2\varepsilon_{\rho\chi\mu\nu}D^{*\rho}\mathop{\partial_\gamma}\limits^{\leftrightarrow}\mathop{\partial^\gamma}\limits^{\leftrightarrow}\bar{D})\nonumber\\&-ie\frac{2}{3\sqrt{35}}g_{\chi_{c2}DD^* \gamma }  A_b \partial^b \partial^\chi \psi_2^{\mu\nu}(9\bar{D}^{*\sigma} \mathop{\partial_\mu}\limits^{\leftrightarrow} \mathop{\partial^\gamma}\limits^{\leftrightarrow} D\varepsilon_{\sigma\gamma\chi\nu} +5\bar{D}^{*\sigma} \mathop{\partial_\nu}\limits^{\leftrightarrow}\mathop{\partial^\gamma}\limits^{\leftrightarrow}D\varepsilon_{\sigma\gamma\chi\mu}+2\varepsilon_{\sigma\chi\mu\nu}\bar{D}^{*\sigma} \mathop{\partial_\gamma}\limits^{\leftrightarrow}\mathop{\partial^\gamma}\limits^{\leftrightarrow}D) \nonumber\\ &-\frac{2e}{3\sqrt{35}} g_{\chi_{c2}D^*D^* \gamma}  A_a \partial^a  \psi_2^{\mu\nu} (11\bar{D}^*_\sigma \mathop{\partial_\mu}\limits^{\leftrightarrow} \mathop{\partial_\nu}\limits^{\leftrightarrow}D^{*}_\rho g^{\rho\sigma} +5\bar{D}^*_\sigma \mathop{\partial^\rho}\limits^{\leftrightarrow} \mathop{\partial_\nu}\limits^{\leftrightarrow} D^{*}_\rho g^{\sigma}_\mu +5\bar{D}^*_\sigma \mathop{\partial^\sigma}\limits^{\leftrightarrow} \mathop{\partial_\nu}\limits^{\leftrightarrow} D^{*}_\rho g^{\rho}_\mu \nonumber\\& +\bar{D}^*_\sigma \mathop{\partial^\rho}\limits^{\leftrightarrow} \mathop{\partial_\mu}\limits^{\leftrightarrow} D^{*}_\rho g^{\sigma}_\nu+\bar{D}^*_\sigma \mathop{\partial^\sigma}\limits^{\leftrightarrow} \mathop{\partial_\mu}\limits^{\leftrightarrow} D^{*}_\rho g^{\rho}_\nu +2\bar{D}^*_\sigma \mathop{\partial^\gamma}\limits^{\leftrightarrow} \mathop{\partial_\gamma}\limits^{\leftrightarrow} D^{*}_\rho g_\mu^{\sigma}g_\nu^{\rho}+2\bar{D}^*_\sigma \mathop{\partial^\gamma}\limits^{\leftrightarrow} \mathop{\partial_\gamma}\limits^{\leftrightarrow} D^{*}_\rho g_\mu^{\rho}g_\nu^{\sigma}),
     \nonumber\\ & +i\ e g_{\chi_{c2} DD_1 \gamma} \chi_{c2}^{\mu\nu}  A_\mu (D\bar{D}_{1\nu}-\bar{D}  D_{1\nu}).
\end{align}

 \begin{align}\label{A13}
    \mathcal{L}_{DD^{(*)}\chi_{c3}(1F) \gamma}= &
  -e\sqrt{3}g_{\chi_{c3}DD^*\gamma} \chi_{c3}^{\mu\nu\theta}(\bar{D} \mathop{\partial_\mu}\limits^{\leftrightarrow} A_\nu D^*_\theta)  -e\sqrt{3}g_{\chi_{c3}DD^*\gamma} \chi_{c3}^{\mu\nu\theta} (D \mathop{\partial_\mu}\limits^{\leftrightarrow} A_\nu \bar{D}^*_\theta)
    \nonumber\\&-ie\sqrt{3}g_{\chi_{c3}D^*D^*\gamma} A^\tau \chi_{c3}^{\mu\nu\theta}(\bar{D}^{*\sigma} \mathop{\partial_\mu}\limits^{\leftrightarrow} \mathop{\partial^\xi}\limits^{\leftrightarrow}D^{*\rho}) ( g_{\rho\sigma} \varepsilon_{\xi\tau\nu\theta}-g_{\sigma\nu}\varepsilon_{\rho\xi\tau\theta}-g_{\rho\nu} \varepsilon_{\sigma\xi\tau\theta}).
\end{align}

The parameters mentioned above are listed here. The coupling constant of coupling between charmed mesons and charmonia  $g_{(Q^{c\bar{c}} D^{(*)}D^{(*)})}$ are determined by matching the theoretical decay width calculated by QPC model. The results are:

\begin{align}\label{A16}
      g_{(\eta_{c2}(2D) DD^*)}&= 1.700,\ g_{(\eta_{c2}(2D) D^*D^*)}= 0.490 \ \text{GeV}^{-1},\ 
      g_{(\chi_{c2}(1F) DD)}=0.934\ \text{GeV}^{-1},\ g_{(\chi_{c2}(1F) DD^*)}=2.710, \nonumber\\ g_{(\chi_{c2}(1F) D^*D^*)}&=1.406\ \text{GeV}^{-1},\ 
      g_{(\chi_{c3}(1F) DD^*)}=5.100\ \text{GeV}^{-1},\ g_{(\chi_{c3}(1F) D^*D^*)} =0.800.
\end{align} 

And the coupling constant of interactions between charmed mesons and $\psi (4230)$ or photon are from the ref. \cite{Peng:2024xui,Bai:2024lps,Li:2021jjt,Chen:2015bma,Gao:2024qth} relatively:

\begin{align}\label{A17}
     g_{\psi DD}&=0.765,\ g_{\psi DD^*}=0.054\ \text{GeV}^{-1},\ g_{\psi D^*D^*}= 1.320, \ 
     g_{D^0 D^{*0}\gamma}= 2.000\ \text{GeV}^{-1}, \ g_{D^+ D^{*-}\gamma}=-0.500\ \text{GeV}^{-1}, \nonumber\\ g_{D^0 D^{0}_1\gamma}&= 1.446\ \text{GeV}^{-1/2}, \ g_{D_1^{+} D^-\gamma}=0.466\ \text{GeV}^{-1/2},\  g_{D^{0}_1 \bar{D}^{*0} \gamma}= 0.610, \ g_{D_1^{+} D^{*-}\gamma}=0.200.
\end{align}

\section{Feynman rules } \label{FL}
In this Appendix, the Feynman rules for the interaction vertexes involved in our calculation are listed here:

\begin{align}\label{B01}
    \braket{D_{(1)}^{(*)}D^{(*)}|\psi}=& g_{\psi DD} \epsilon^{l}_{\psi}(q_{1l}-q_{2l}) +g_{\psi DD^*} \varepsilon^{hl\alpha j} \epsilon_{\psi j} \epsilon^*_{D^* \alpha} p_{1h}(q_{2l}-q_{1l}) -g_{\psi DD^*} \varepsilon^{lj\lambda h} \epsilon_{\psi j} \epsilon^*_{\bar{D}^* \lambda} p_{1l}(q_{2h}-q_{h}) \nonumber\\ & +g_{\psi D^*D^*}  \epsilon^{l}_{\psi} \epsilon_{D^*}^{*\alpha} \epsilon_{\bar{D}^*}^{*\lambda} [g_{\alpha l} q_{1 \lambda}-q_{2\alpha}g_{\lambda l}+g_{\alpha\lambda}(q_{2l}-q_{1l})] +i g_{\psi DD_1} \epsilon_\psi^\alpha \epsilon^*_{D_1 \alpha}
\end{align} 

\begin{align}\label{B02}
    \braket{\gamma |D_{(1)}^{(*)}D^{(*)}}&= -e \epsilon_\gamma^{*m}(q_m+q_{1m}) -\frac{e f_{DD^*\gamma}}{4} \varepsilon^{m n k t} \epsilon_\gamma^{*r} \epsilon_{D^*}^{*\tau} (p_{2m}g_{nr}-p_{2n}g_{mr})(q_k g_{\tau t}- q_t g_{\tau k})\nonumber\\ & +\frac{e f_{DD^*\gamma}}{4} \varepsilon^{m n k t} \epsilon_\gamma^{*r} \epsilon_{D^*}^{\beta} (p_{2m}g_{nr}-p_{2n}g_{mr})(q_{1k} g_{\beta t}- q_{1t} g_{\beta k}) -e\epsilon_\gamma^{*m} \epsilon_{D^*}^{\beta} \epsilon_{D^*}^{*\tau} (q_\beta g_{m \tau}+q_{1\tau}g_{m \beta}+g_{\beta \tau}(q_m+q_{1m}))\nonumber\\ & +ief_{DD_1\gamma} \epsilon_\gamma^{*\beta}\epsilon_{D_1\beta}+ie f_{D^*D_1\gamma} \varepsilon^{m n \tau \beta} \epsilon_\gamma^{*r} \epsilon_{D_1\beta}\epsilon^*_{D^*\tau}(p_{2m}g_{nr}-p_{2n}g_{mr})
\end{align}

\begin{align}\label{B04}
    \braket{\eta_{c2}|D^{*}D^{(*)}}=&-2g_{\eta_{c2}DD^*}g_{\nu\rho}(q_{2\mu}-q_{\mu})\epsilon_{\eta2}^{*\mu\nu}\epsilon_{D^*}^{\rho} -2g_{\eta_{c2}DD^*}g_{\sigma\nu}(q_{2\mu}-q_{\mu})\epsilon_{\eta2}^{*\mu\nu}\epsilon_{\bar{D}^*}^{\sigma}
    \nonumber\\ & -2g_{\eta_{c2}D^*D^*} \epsilon_{\eta2}^{*\mu\nu}\epsilon_{\bar{D}^*}^{\sigma} \epsilon_{D^*}^{\rho}p_3^\theta  \varepsilon_{\rho\sigma\theta\nu}(q_{\mu}-q_{2\mu})
\end{align} 

\begin{align}\label{B08}
    \braket{\chi_{c2}|D^{(*)}D^{(*)}}=& \frac{2i\sqrt{7}}{5} g_{\chi_{c2} DD}(q_{\mu}q_{2\nu}+q_{\nu}q_{2\mu}-q_{\mu}q_{\nu}-q_{2\mu}q_{2\nu}) \epsilon_{\chi_{c2}}^{*\mu\nu} \nonumber\\ 
    & +\frac{2i}{3\sqrt{35}} g_{\chi_{c2} DD^*} p_{3}^a [2\varepsilon_{\rho a \mu\nu}(q_{2}^b-q^{b})(q_{2b}-q_{b}) +5\varepsilon_{\rho b a\mu}(q^b-q_{2}^b)(q_{\nu}-q_{2\nu}) +9\varepsilon_{\rho b a\nu }(q^b-q_{2}^b)(q_{\mu}-q_{2\mu})]\epsilon_{\chi_{c2}}^{*\mu\nu}\epsilon^{\rho}_{D^*}  \nonumber\\ 
    & +\frac{2i}{3\sqrt{35}} g_{\chi_{c2} DD^*} p_{3}^a [2\varepsilon_{\sigma a \mu\nu}(q_{2}^b-q^{b})(q_{2b}-q_{b})+5\varepsilon_{\sigma b a\mu}(q^b-q_{2}^b)(q_{\nu}-q_{2\nu}) +9\varepsilon_{\sigma b a\nu }(q^b-q_{2}^b)(q_{\mu}-q_{2\mu})]\epsilon_{\chi_{c2}}^{*\mu\nu}\epsilon^{\sigma}_{\bar{D}^*} \nonumber\\ 
    & -\frac{2i}{3\sqrt{35}}g_{\chi_{c2} D^*D^*} \epsilon_{\chi_{c2}\mu\nu}^{*} \epsilon_{D^* \rho}\epsilon_{\bar{D}^*\sigma} [(2g^{\mu\sigma}g^{\nu\rho}+2g^{\mu\rho} g^{\nu\sigma})(q_2^n-q^n)(q_{2n}-q_n)  +5g^{\mu \rho} (q_2^\nu-q^\nu)(q_{2}^{\sigma}-q^\sigma)  \nonumber\\
    & +5g^{\mu \sigma} (q_2^\nu-q^\nu)(q_2^{\rho}-q^\rho) +g^{\nu\rho} (q_2^\mu-q^\mu)(q_{2}^{\sigma}-q^\sigma)  +g^{\nu\sigma} (q_2^\mu-q^\mu)(q_{2}^{\rho}-q^\rho) +11 g^{\rho\sigma} (q_2^\mu-q^\mu)(q^{\nu}_{2}-q^\nu)]
\end{align}

\begin{align}\label{B09}
    \braket{\chi_{c3}|D^{*}D^{(*)}}=& -i\sqrt{3} g_{\chi_{c3}DD^*} g_{\rho\theta}(q_{1\mu}-q_{2\mu})(q_{1\nu}-q_{2\nu})\epsilon_{\chi_{c3}}^{*\mu\nu\theta}\epsilon_{D^*}^{\rho} -i\sqrt{3} g_{\chi_{c3}DD^*} g_{\sigma\theta}(q_{2\mu}-q_{1\mu})(q_{2\nu}-q_{1\nu})\epsilon_{\chi_{c3}}^{*\mu\nu\theta}\epsilon_{\bar{D}^*}^{\sigma}
    \nonumber\\ &  +i\sqrt{3} g_{\chi_{c3}D^*D^*} p_{3}^a[(q_{\mu}-q_{2\mu})(q_{}^{\gamma}-q_{2}^{\gamma})]  (g_{\sigma\rho}\varepsilon_{\gamma a \nu \theta}-g_{\nu\rho}\varepsilon_{\sigma \gamma a \theta} -g_{\nu\sigma}\varepsilon_{\rho \gamma a \theta}) \epsilon_{\chi_{c3}}^{*\mu\nu\theta}\epsilon_{D^*}^{\rho}\epsilon_{\bar{D}^*}^{\sigma}.
\end{align}

\section{Gauge invariance} \label{GI}

Notice that the triangle diagrams include the $DD^*\gamma$ vertices which satisfy Ward Takahashi identity automatically. The other amplitudes of $\eta_2(2D)$ can be expanded after calculating the loop integral as listed here

\begin{align}
\sum\mathcal{M}_{\text{tri}}^{\eta_{c2}}=& A p_{2}^\mu \varepsilon^{vabc}p_{3a} \epsilon^*_{\gamma b} \epsilon_{\psi c}\epsilon^*_{\eta_{c2} \mu\nu} +B p_{2}^\mu \varepsilon^{vabc}p_{2a} \epsilon^*_{\gamma b} \epsilon_{\psi c} \epsilon^*_{\eta_{c2} \mu\nu} + C p_{2}^\mu p_{2}^\nu \varepsilon^{abcd} p_{2a} p_{3b} \epsilon^*_{\gamma c}\epsilon_{\psi d}\epsilon^*_{\eta_{c2} \mu\nu} \nonumber\\ & + D \epsilon^\mu_{\psi} \varepsilon^{vabc}p_{2a}  p_{3b} \epsilon^*_{\gamma c} \epsilon^*_{\eta_{c2} \mu\nu} +E \epsilon^{*\mu}_{\gamma} \varepsilon^{vabc}p_{2a}  p_{3b} \epsilon_{\psi c} \epsilon^*_{\eta_{c2} \mu\nu} \nonumber\\ & + F p_2^\mu (p_3^d \epsilon^{*}_{\gamma d}) \varepsilon^{vabc} p_{2a}  p_{3b} \epsilon_{\psi c} \epsilon^*_{\eta_{c2} \mu\nu}  +  G p_2^\mu (p_2^d \epsilon_{\psi d}+p_3^d \epsilon_{\psi d}) \varepsilon^{vabc} p_{2a}  p_{3b} \epsilon^*_{\gamma c} \epsilon^*_{\eta_{c2} \mu\nu},
\nonumber\\ \sum\mathcal{M}^{\eta_{c2}}_{\text{con}} =& H p_{2}^\mu \varepsilon^{vabc}p_{2a} \epsilon^*_{\gamma b} \epsilon_{\psi c} \epsilon^*_{\eta_{c2} \mu\nu} + H p_{2}^\mu \varepsilon^{vabc}p_{3a} \epsilon^*_{\gamma b} \epsilon_{\psi c}\epsilon^*_{\eta_{c2} \mu\nu}.
\end{align}

It is evident that the $B$, $C$, $D$, and $F$ terms satisfy gauge invariance, whereas the remaining terms do not. Replacing $\epsilon^*_{\gamma\mu}$ with $p_{2\mu}$ yields the relation between the contracted terms and the triangle terms

\begin{align}
    H=&-[A+E+G (p2 \cdot p3)], \\
    \sum\mathcal{M}_{con} =& -[A+E+ G (p2 \cdot p3)] (p_{2}^\mu \varepsilon^{vabc}p_{2a} \epsilon^*_{\gamma b} \epsilon_{\psi c} \epsilon^*_{\eta_{c2} \mu\nu} + p_{2}^\mu \varepsilon^{vabc}p_{3a} \epsilon^*_{\gamma b} \epsilon_{\psi c}\epsilon^*_{\eta_{c2} \mu\nu}).
\end{align}

Other amplitudes can be calculated in the same way. The amplitudes of the $\chi_{c2} (1F)$ state can be calculated as:

\begin{align}
\sum\mathcal{M}_{\text{tri}}^{\chi_{c2}}=& A \epsilon^*_{\gamma \mu} \epsilon_{\psi \nu}\epsilon_{\chi_{c2}}^{*\mu\nu} +B \epsilon^*_{\gamma \nu}  p_{2\mu}  (\epsilon_{\psi a} p_2^a ) \epsilon_{\chi_{c2}}^{*\mu\nu} +C p_{2\nu}  p_{2\mu}  (\epsilon_{\psi a} \epsilon^*_{\gamma a}) \epsilon_{\chi_{c2}}^{*\mu\nu} + D p_{2\mu} \epsilon_{\psi \nu} ( \epsilon^*_{\gamma a} p_3^a ) \epsilon_{\chi_{c2}}^{*\mu\nu} \nonumber\\ 
&  + E p_{2\mu} \epsilon^*_{\gamma \nu} (\epsilon_{\psi a} p_3^a ) \epsilon_{\chi_{c2}}^{*\mu\nu} +F p_{2\nu}  p_{2\mu}  (\epsilon_{\psi a} p_2^a )( \epsilon^*_{\gamma b} p_3^b) \epsilon_{\chi_{c2}}^{*\mu\nu}    +G p_{2\nu}  p_{2\mu}  (\epsilon_{\psi a} p_3^a )( \epsilon^*_{\gamma b} p_3^b) \epsilon_{\chi_{c2}}^{*\mu\nu}, 
\nonumber\\ 
\sum\mathcal{M}^{\chi_{c2}}_{\text{con}} =& H p_{2\nu}  p_{2\mu} ( \epsilon^*_{\gamma b} p_3^b) \epsilon_{\chi_{c2}}^{*\mu\nu} +I p_{2\nu}  p_{2\mu}  (\epsilon_{\psi a} p_2^a )( \epsilon^*_{\gamma b} p_3^b) + J p_{2\mu} \epsilon_{\psi \nu} ( \epsilon^*_{\gamma a} p_3^a ).
\end{align}

  where we do not consider the triangle diagrams include the $DD^* \gamma$ vertices because they satisfy Ward Takahashi identity automatically. And we can obtain similar relation between the contract terms and the triangle terms as

\begin{align}
    (p2\cdot p3)H=&-[A+D (p2 \cdot p3)], \ \ 
    (p2\cdot p3)J=-[E+G (p3 \cdot p2)],  \ \ 
    (p2\cdot p3)J=-[B+F (p3 \cdot p2)], \\
   \sum\mathcal{M}^{\chi_{c2}}_{\text{con}} =& -[A+D (p2 \cdot p3)]/(p2\cdot p3)  p_{2\nu}  p_{2\mu} \epsilon_{\chi_{c2}}^{*\mu\nu}  -[E+G (p3 \cdot p2)]/ (p2\cdot p3)  p_{2\nu}  p_{2\mu}  (\epsilon_{\psi a} p_2^a )  \epsilon_{\chi_{c2}}^{*\mu\nu} \nonumber\\ & -[B+F (p3 \cdot p2)]/ (p2\cdot p3) p_{2\mu} \epsilon_{\psi \nu}  (\epsilon^*_{\gamma a} p_3^a )\epsilon_{\chi_{c2}}^{*\mu\nu}  .
\end{align}

Using the same method, the $\chi_{c3} (1F)$ state can be calculated as:

\begin{align}
\sum\mathcal{M}_{\text{tri}}^{\chi_{c3}}=& A p_{2}^\mu p_2^\nu \varepsilon^{\theta abc}p_{3a} \epsilon^*_{\gamma b} \epsilon_{\psi c}\epsilon^*_{\chi_{c3 \mu\nu\theta}} + B p_{2}^\mu p_2^\nu \varepsilon^{\theta abc} p_{2a} \epsilon^*_{\gamma b} \epsilon_{\psi c}\epsilon^*_{\chi_{c3\mu\nu\theta}} +C p_{2}^\mu p_2^\nu p_2^\theta \varepsilon^{abcd} p_{2a}  p_{3b} \epsilon^*_{\gamma c} \epsilon_{\psi d}\epsilon^*_{\chi_{c3\mu\nu\theta}} \nonumber \\ 
& + D p_{2}^\mu \epsilon_{\psi}^{\nu} \varepsilon^{\theta abc}p_{2a} p_{3b} \epsilon^*_{\gamma c} \epsilon^*_{\chi_{c3\mu\nu\theta}} +E p_{2}^\mu \epsilon^{*\nu}_{\gamma} \varepsilon^{\theta abc}p_{2a} p_{3b} \epsilon_{\psi c}  \epsilon^*_{\chi_{c3 \mu\nu\theta}} + F p_{2}^\mu p_2^\nu p_{2e} \epsilon_{\psi}^{e} \varepsilon^{\theta abc}p_{2a} p_{3b} \epsilon^*_{\gamma c} \epsilon^*_{\chi_{c3\mu\nu\theta}} \nonumber\\
& + G p_{2}^\mu p_2^\nu p_{3e} \epsilon_{\psi}^{e} \varepsilon^{\theta abc}p_{2a} p_{3b} \epsilon^*_{\gamma c} \epsilon^*_{\chi_{c3\mu\nu\theta}} +H p_{2}^\mu p_2^\nu \epsilon^{*e}_{\gamma} p_{3e} \varepsilon^{\theta abc}p_{2a} p_{3b} \epsilon_{\psi c}  \epsilon^*_{\chi_{c3 \mu\nu\theta}},
\nonumber\\ \sum\mathcal{M}_{\text{con}}^{\chi_{c3}}=& I p_{2}^\mu p_2^\nu \varepsilon^{\theta abc}p_{3a} \epsilon^*_{\gamma b} \epsilon_{\psi c}\epsilon^*_{\chi_{c3\mu\nu\theta}} + I p_{2}^\mu p_2^\nu \varepsilon^{\theta abc} p_{2a} \epsilon^*_{\gamma b} \epsilon_{\psi c}\epsilon^*_{\chi_{c3 \mu\nu\theta}}.
\end{align}

Then we have the relations between the contract terms and the triangle terms as

\begin{align}
    I=&-[A+E+H(p2 \cdot p3)], \\
   \sum\mathcal{M}^{\chi_{c3}}_{\text{con}} =& -[A+E+H(p2 \cdot p3)] p_{2}^\mu p_2^\nu \varepsilon^{\theta abc}p_{3a} \epsilon^*_{\gamma b} \epsilon_{\psi c}\epsilon^*_{\chi_{c3\mu\nu\theta}} -[A+E+H(p2 \cdot p3)] p_{2}^\mu p_2^\nu \varepsilon^{\theta abc} p_{2a} \epsilon^*_{\gamma b} \epsilon_{\psi c}\epsilon^*_{\chi_{c3 \mu\nu\theta}}.
\end{align}

\end{widetext}

\end{appendix}

\bibliography{ref.bib}

\end{document}